\pdfoutput=1
\documentclass[11pt]{article}

\usepackage[preprint]{acl}

\usepackage{times}
\usepackage{latexsym}

\usepackage[T1]{fontenc}
\usepackage[utf8]{inputenc}

\usepackage{microtype}

\usepackage{inconsolata}

\usepackage{pifont}

\usepackage{hyperref}       % hyperlinks
\usepackage{url}            % simple URL typesetting
\usepackage{booktabs}       % professional-quality tables
\usepackage{amsfonts}       % blackboard math symbols
\usepackage{nicefrac}       % compact symbols for 1/2, etc.
\usepackage{xcolor}         % colors
\usepackage{paralist}
\usepackage{mathrsfs}
\usepackage{multirow}
\usepackage{pgfplots}
\usepackage{scalefnt}
\usepackage{amsthm,amsmath,amssymb}
\usepackage{mathtools}
\usepackage{bm}
\usepackage{colortbl}
\usepackage{array}
\usepackage{stfloats}
\usepackage{tikz-dependency}
\usepackage{tikz}
\usepackage{siunitx}
\usepackage{graphicx}
\usepackage[ruled]{algorithm2e}
\SetAlFnt{\scriptsize}
\SetAlCapFnt{\scriptsize}
\SetAlCapNameFnt{\scriptsize}
\usepackage{lineno}
\usepackage{subfigure}
\usepackage{helvet}
\usepackage{makecell}
\usepackage{enumitem}
\usepackage{wrapfig}
\usepackage{tcolorbox}
\usepackage{courier}
\usepackage{caption}
\usepackage{CJKutf8}

\usepackage{float}

\title{Beyond Chunking: Discourse-Aware Hierarchical Retrieval for Long Document Question Answering}

\author{
  \textbf{Huiyao Chen\textsuperscript{1,2},} 
  \textbf{Yi Yang\textsuperscript{1},} 
  \textbf{Yinghui Li\textsuperscript{1},}
  \textbf{Meishan Zhang\textsuperscript{1},}$\thanks{~~Corresponding author: Meishan Zhang.}$
  \textbf{Baotian Hu\textsuperscript{1,2},}
  \textbf{Min Zhang\textsuperscript{1,2}}
  \\
  \textsuperscript{1}Institute of Computing and Intelligence, Harbin Institute of Technology (Shenzhen), China
  \\
  \textsuperscript{2} Shenzhen Loop Area Institute (SLAI)
  \\
  \texttt{\{chenhy1018,yangi10010730,mason.zms\}@gmail.com}
}

\pgfplotsset{compat=1.18} 
\begin{document}
\maketitle
\begin{abstract}
Existing long-document question answering systems typically process texts as flat sequences or use heuristic chunking, which overlook the discourse structures that naturally guide human comprehension.
We present a discourse-aware hierarchical framework that leverages rhetorical structure theory (RST) for long document question answering.
Our approach converts discourse trees into sentence-level representations and employs LLM-enhanced node representations to bridge structural and semantic information.
The framework involves three key innovations: language-universal discourse parsing for lengthy documents, LLM-based enhancement of discourse relation nodes, and structure-guided hierarchical retrieval.
Extensive experiments on four datasets demonstrate consistent improvements over existing approaches through the incorporation of discourse structure, across multiple genres and languages. Moreover, the proposed framework exhibits strong robustness across diverse document types and linguistic settings.

\end{abstract}

\section{Introduction}

Document question answering represents a fundamental challenge in natural language processing, with applications spanning from academic research to enterprise knowledge management \citep{chen-etal-2017-reading,karpukhin-etal-2020-dense}.
To date, large language models (LLMs) have achieved remarkable success on the task, particularly for short documents such as SQuAD \citep{rajpurkar-etal-2016-squad}, which has an average context length of 117 words \citep{coqa-2019}, reaching human-level performance with F1 scores exceeding 85\% \citep{chowdhery2023palm,malladi2023fine}.
However, as document length increases, their performance degrades significantly.
For example, on challenging long document datasets such as QASPER,  state-of-the-art LLM models only achieve performance with less than 50\% F1 scores \citep{shaham-etal-2022-scrolls,bai-etal-2024-longbench}.
This performance gap highlights long document question answering as a critical and underexplored research frontier.

\begin{figure}[t]
    \centering
    \includegraphics[width=1.0\linewidth]{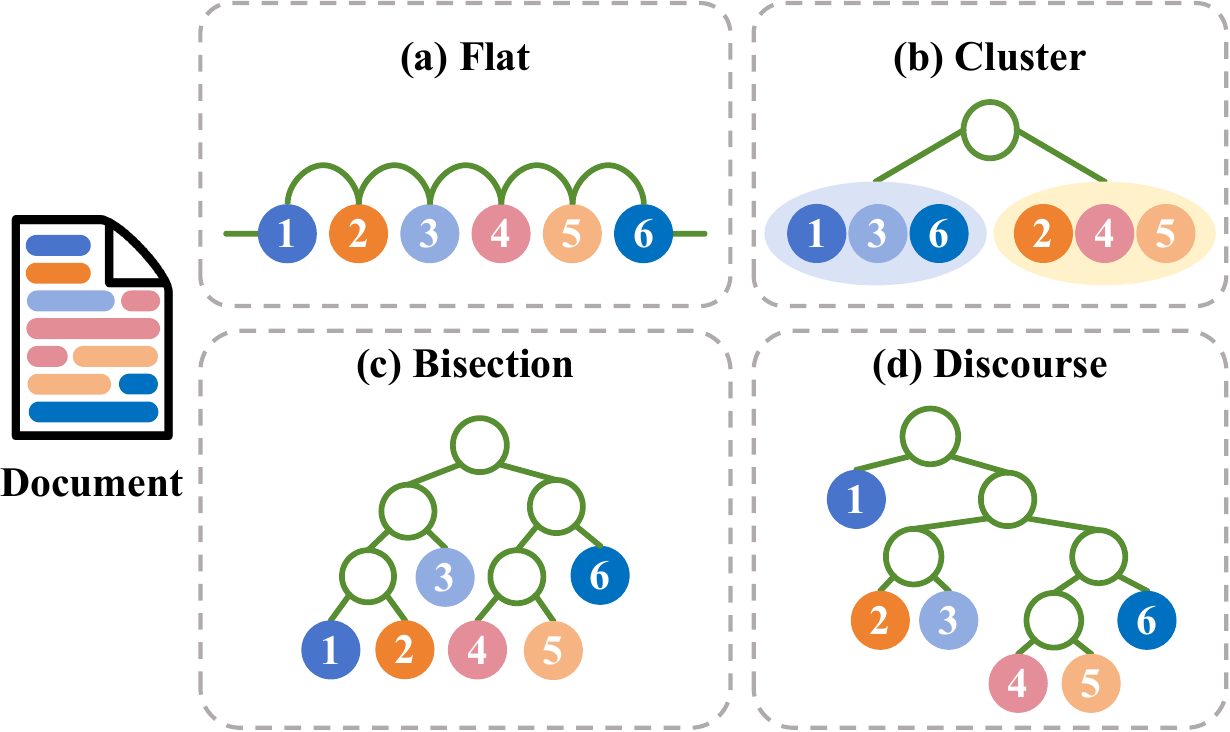}
    \caption{Comparison of document modeling approaches for long document QA. Numbers (1-6) show sentence order in original document, with similar colors indicating semantic relationships. Four approaches are compared: (a) Flat sequential modeling, (b) Bottom-up semantic clustering, (c) Bisection-based adjacent grouping, and (d) Our discourse-aware approach that preserves both semantic and discourse structures.}
    \label{fig:intro}
\end{figure}

The most straightforward approach to handling long documents involves flat sequential modeling, where texts are processed as linear sequences or divided into fixed-size chunks \citep{lewis2020retrieval,DBLP:journals/corr/abs-2002-08909}.
This chunking-based paradigm has gained widespread adoption due to its simplicity and computational efficiency, enabling scalable processing of large document collections.
Building upon this foundation, recent advances have introduced tree-based structures to better capture document organization and move beyond simple chunking.
These innovations include semantic clustering methods like RAPTOR \citep{DBLP:conf/iclr/SarthiATKGM24} that recursively group similar content, and bisection-based approaches that maintain local coherence through adjacent segment grouping \citep{liu-lapata-2019-hierarchical,karpukhin-etal-2020-dense}.
These hierarchical methods represent significant improvements by capturing semantic similarity and textual coherence, respectively, offering more sophisticated document modeling capabilities.

Discourse structure provides a principled linguistic foundation for document organization that goes beyond surface-level similarities.
Consider a document where the first sentence serves as a topic statement, followed by contrasting ideas (sentences 2-3), parallel supporting evidence (sentences 4-5), and a concluding statement (sentence 6).
These discourse relationships naturally capture how humans organize and comprehend information in documents, providing guidance for more semantically appropriate chunking and hierarchical organization in retrieval.
As a mature theoretical framework, discourse analysis has demonstrated effectiveness across various NLP tasks \citep{feng-hirst-2014-linear,cohan-etal-2018-discourse,sogaard-etal-2021-need}, presenting an opportunity to develop more principled retrieval methods that align with human document comprehension patterns.

Therefore, in this work, we present \textbf{DISRetrieval} (\textbf{DIS}course-aware hierarchical \textbf{Retrieval}), the first systematic approach that leverages rhetorical structure theory (RST) to enhance long document question answering through discourse-aware modeling with cross-lingual applicability.
Our approach tackles three key technical challenges: First, we introduce RST adaptations along two dimensions: granularity adaptation that shifts processing to the sentence level for efficiency, and language adaptation that enables cross-lingual applicability through LLM-based data augmentation.
Second, we introduce LLM-based node enhancement that enriches intermediate nodes with both discourse structure and semantic content.
Third, we design structure-guided evidence retrieval mechanisms that leverage discourse organization for effective information extraction.

Comprehensive experiments across four challenging benchmarks:
QASPER \citep{qasper-dataset} for research paper understanding, QuALITY \citep{quality-dataset} for reading comprehension, NarrativeQA \cite{narrativeqa-dataset} for book-length narrative analysis and MultiFieldQA-zh \citep{bai-etal-2024-longbench} for Chinese documents question answering demonstrate substantial improvements over existing methods.
Detailed ablation studies validate the effectiveness of our discourse-aware modeling and structure-guided processing, showing consistent gains across diverse scenarios.
Our framework successfully captures both fine-grained semantic details and document-level organizational patterns, providing a robust solution for discourse-informed question answering.
Our main contributions can be summarized as follows:
\begin{compactitem}
\item A novel discourse-aware hierarchical framework with granularity and language adaptations enabling efficient cross-lingual long document QA.
\item An innovative LLM-enhanced hierarchical retrieval mechanism enabling multi-granularity evidence selection.
\item Comprehensive empirical validation across diverse datasets, architectures, and languages.
\end{compactitem}
Our code and datasets are publicly available at \href{https://github.com/DreamH1gh/DISRetrieval}{github/DreamH1gh/DISRetrieval} to facilitate future research.

\section{Related Work}

Recent advances in LLMs have demonstrated impressive capabilities across diverse text comprehension tasks \citep{brown2020language,chowdhery2023palm}.
However, extended documents present challenges due to context length constraints and computational complexity \citep{DBLP:conf/iclr/Tay0ASBPRYRM21,zaheer2020big,beltagy2020longformer,ding2023longnet,ainslie-etal-2023-gqa}.
Traditional approaches operate on short segments without considering broader context \citep{liu-lapata-2019-hierarchical,guo-etal-2022-longt5,DBLP:conf/iclr/RaePJHL20}, while recent innovations explore chunking-free extraction \citep{zhao-etal-2024-seer} and in-context retrieval \citep{qian-etal-2024-grounding,wang-etal-2023-simlm,DBLP:journals/jmlr/IzacardLLHPSDJRG23,DBLP:conf/iclr/0002IWXJ000023}.

Retrieval-based methods segment documents before retrieving relevant evidence \citep{lewis2020retrieval,karpukhin-etal-2020-dense,izacard2020leveraging}.
These have evolved from flat segmentation \citep{guu2020retrieval} and early hierarchical approaches \citep{tang2017qalink} to sophisticated methods including semantic clustering \citep{DBLP:conf/iclr/SarthiATKGM24}, bisection-based techniques \citep{DBLP:conf/icml/ZhangZSL20,ivgi2023efficient}, hybrid systems \citep{liu-etal-2021-dense-hierarchical,arivazhagan-etal-2023-hybrid}, and structure conversion approaches \citep{longrefiner}.
Discourse structure has been explored for QA through discourse-based systems \citep{santhosh2012discourse}, long-form answer analysis \citep{xu-etal-2022-answer}, and structure-discourse graphs \citep{nair2023drilling,du-etal-2023-structure}.
However, existing approaches either rely on surface-level semantic similarity for retrieval or apply discourse primarily to answer generation tasks, overlooking linguistically-informed discourse structures for evidence retrieval \citep{DBLP:journals/corr/abs-1907-11692,wolf-etal-2020-transformers}.

\section{Method}

Discourse analysis provides systematic frameworks for understanding how textual segments relate beyond surface-level semantics \cite{mann1988rhetorical, carlson-etal-2001-building}.
RST formalizes this by representing documents as hierarchical trees where leaf nodes correspond to elementary discourse units (EDUs) and internal nodes encode rhetorical relations such as contrast, elaboration, and summary \cite{marcu2000theory}.
Recent advances in neural discourse parsing \cite{yu2018transition, kobayashi2020top,yu2022rst,yu-etal-2022-rst} have enabled automatic RST construction at scale \cite{maekawa-etal-2024-obtain,yuan2025cross}.

\begin{figure*}[t]
    \centering
    \includegraphics[width=1.0\linewidth]{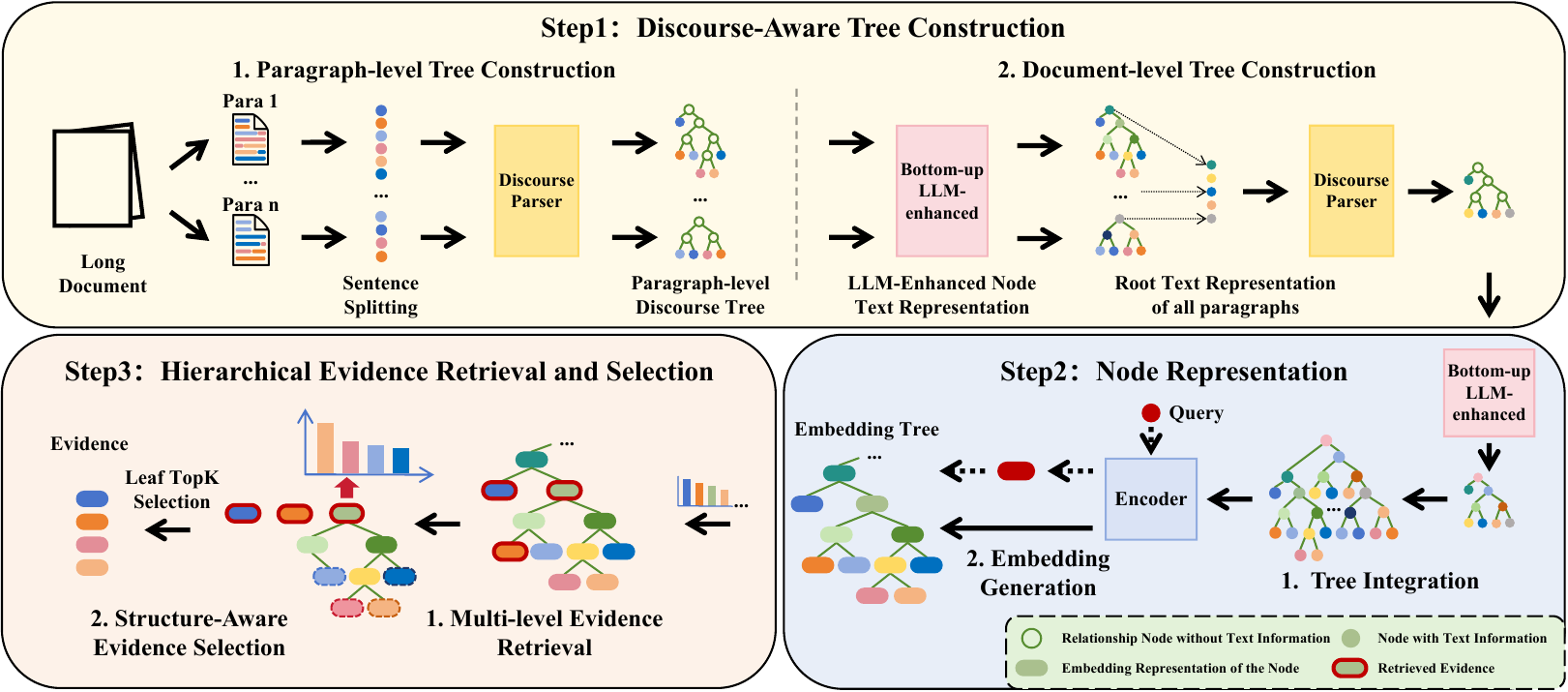}
    \caption{Overview of the DISRetrieval framework consisting of three main steps: (1) Discourse-Aware Tree Construction through paragraph-level parsing (left) and document-level integration (center), (2) Node Representation via tree integration and dense encoding, and (3) Hierarchical Evidence Retrieval with multi-level selection. Color coding indicates discourse relations.}
    \label{fig:framework}
\end{figure*}

To address the challenge of bridging between discourse theory and document retrieval, we propose \textbf{DISRetrieval}, a discourse-aware framework that systematically incorporates RST structure into document retrieval as shown in Figure~\ref{fig:framework}.
Our framework operates in three stages: (1) constructing hierarchical discourse trees through sentence-level RST parsing, (2) enriching tree nodes with semantic representations via LLM enhancement and dense encoding, and (3) performing structure-guided evidence retrieval.
Figure~\ref{fig:llm_enhancement} provides a concrete example showing how a research paper paragraph is transformed into a hierarchical discourse structure.

\subsection{Discourse-Aware Tree Construction}
\label{Discourse_Guided}

We construct a hierarchical tree structure that captures semantic content and organizational structure of documents through discourse analysis.

\subsubsection{RST Adaptation}
\label{section:rst_adaptation}
To effectively apply RST to long document question answering, we introduce two critical adaptations that address computational efficiency and cross-lingual applicability.

\paragraph{Granularity Adaptation.}
Traditional RST operates on fine-grained EDUs, creating computational overhead and semantic fragmentation for long documents.
We address this challenge by shifting RST processing to the sentence level, achieving better efficiency while maintaining semantic coherence.
Specifically, we train a sentence-level parser by converting existing EDU-based datasets through two operations: (1) merging intra-sentence EDUs into unified sentence units, and (2) determining inter-sentence relationships via lowest common ancestor analysis in the original EDU-level discourse trees.
This enables our parser to capture meaningful discourse relations efficiently.

\paragraph{Language Adaptation.}
To enable cross-lingual applicability beyond English, we develop a language-universal discourse parser through LLM-based multilingual data augmentation.
We employ GPT-4o to translate the RST-DT training corpus into target languages while preserving discourse structures at the sentence level.
The translated data is combined with the original corpus to train a unified parser $f_\text{discourse}$ that generalizes across languages without requiring language-specific annotations.
Complete implementation details are provided in Appendix~\ref{app:discourse_parser} and \ref{RST-DT_Translation}.

\subsubsection{Complete Tree Construction Process}
\label{section:tree_construction}

Building upon our adapted sentence-level discourse parser, we develop a hierarchical discourse modeling framework that constructs document-level discourse trees through a two-phase process.

\paragraph{Phase 1. Paragraph-Level Tree Construction.}
We transform each paragraph's sentence sequence $\boldsymbol{S_i} = \{s_{i,1}, s_{i,2}, ..., s_{i,m}\}$ into a local discourse tree $T_i$ using $f_\text{discourse}$.
Each tree captures sentence connections within the paragraph, preserving fine-grained intra-paragraph discourse structure.

\paragraph{Phase 2. Document-Level Tree Construction.}
We construct global document structure through bottom-up LLM enhancement.
For each internal node $v$ in paragraph-level trees, we apply adaptive processing:
\begin{equation}
    v^* = \begin{cases}
        f_\text{LLM}(v_l, v_r), & \text{if } |v_l| + |v_r| \geqslant \tau \\
        f_\text{merge}(v). & \text{otherwise}
    \end{cases}
\label{eq:LLM_enhance}
\end{equation}
When combined child length exceeds threshold $\tau$, $f_\text{LLM}$ generates concise summaries. Otherwise, $f_\text{merge}$ directly concatenates content.
Figure~\ref{fig:llm_enhancement} illustrates how rhetorical relations transform into semantic representations.

The root representations from paragraph trees serve as input for document-level tree construction.
We apply the same RST parser to these summaries, creating a hierarchical structure capturing document-wide discourse relationships.

\subsection{Discourse-Aware Node Representation}
\label{node_representation}

A critical challenge in discourse-based retrieval lies in bridging the semantic gap between abstract rhetorical relations and concrete textual content required for neural retrieval.
We develop a semantic enhancement framework through three key stages.

\paragraph{Bottom-up Semantic Enhancement.}
The document-level tree $T_\text{doc}$ initially lacks concrete semantic content at internal nodes.
We apply the same bottom-up LLM enhancement strategy (Equation~\ref{eq:LLM_enhance}) to generate meaningful textual representations for these structural nodes.
Starting from leaf nodes and moving upward, this process systematically transforms each internal node from abstract structural placeholders into semantically rich representations that capture the combined content of their subtrees.
The result is a fully enhanced tree where all nodes possess concrete textual content suitable for semantic matching.

\paragraph{Multi-Level Tree Integration.}
We implement a substitution-based integration mechanism that combines the enhanced document-level tree $T_\text{doc}^*$ with paragraph-level trees $T_i$ into a unified discourse structure.
Each leaf node in $T_\text{doc}^*$ is systematically replaced with its corresponding paragraph-level discourse tree:
\begin{equation}
T_\text{D} = T_\text{doc}^*[l_i \leftarrow T_i, \forall i \in \{1, 2, ..., n\}].
\label{eq:tree_integration}
\end{equation}

This integration enables multi-granularity discourse modeling, where fine-grained intra-paragraph sentence relationships coexist with coarse-grained inter-paragraph discourse connections within a single hierarchical structure.

\paragraph{Node Encoding.}
\label{eq:embedding}
To enable efficient retrieval, we transform the integrated discourse tree $T_\text{D}$ into dense embeddings using a pre-trained encoder:
\begin{equation}
\mathbf{e}_v = f_\text{enc}(v), \quad \forall v \in \mathcal{N}(T_\text{D}).
\label{eq:node_encoding}
\end{equation}
The resulting embedding tree $\mathcal{T}_\text{emb} = (T_\text{D}, \{\mathbf{e}_v\})$ preserves both the hierarchical discourse structure and semantic information, enabling structure-aware retrieval through dense vector operations.

\subsection{Structure-Guided Evidence Retrieval}
\label{Hierarchical_retreival}

We introduce a structure-aware retrieval mechanism that leverages the hierarchical nature of discourse trees to select evidence at multiple granularity levels. Unlike flat retrieval methods that operate on uniform text segments, our approach exploits discourse structure to implement a dual-selection strategy that balances local semantic relevance with global discourse coherence.

\paragraph{Retrieval Process.}
Given a query $q$, we first compute semantic similarities for all nodes in the embedding tree:
\begin{equation}
\text{score}(v) = \text{cos}(f_\text{enc}(q), \mathbf{e}_v), \forall v \in \mathcal{N}(T_\text{D}).
\label{eq:similarity_scoring}
\end{equation}

\paragraph{Dual-Selection Strategy.}
Nodes are ranked by relevance and processed through two complementary selection mechanisms:

\begin{tcolorbox}[colback=gray!5,colframe=gray!40,boxsep=0pt,top=1pt,bottom=1pt,left=2pt,right=2pt]
    \begin{itemize}[leftmargin=8pt,nosep]
        \item \textbf{Direct leaf selection} for high-relevance sentences
        \item \textbf{Hierarchical expansion} with top-$k$ subtree selection for internal nodes
        \item \textbf{Redundancy elimination} to prevent duplicates
    \end{itemize}
\end{tcolorbox}

For leaf nodes with high relevance, we select them directly. For internal nodes, we perform controlled subtree expansion by selecting the top-$k$ most relevant unused leaves from their subtrees. This ensures both specific sentence-level evidence and coherent discourse segments are captured while eliminating redundancy.

Algorithm~\ref{alg:retrieval} details the complete process, balancing semantic relevance with discourse coherence.

\begin{algorithm}[t]
    \caption{Structure-Guided Evidence Retrieval}
    \label{alg:retrieval}
    
    \KwIn{Query $q$, discourse tree $T$, evidence limit $K$, subtree selection size $k$}
    \KwOut{Evidence set $E$}
    
    \tcc{Compute node similarities and rank}
    $\mathbf{e_q} \gets \text{Encoder}(q)$ \;
    $\text{scores}[v] \gets \text{cosine}(\mathbf{e_q}, \mathbf{e_v})$ for all $v \in T$ \;
    $V_\text{ranked} \gets \text{sort}(\text{scores}, \text{descending})$ \;
    
    \tcc{Structure-aware selection}
    $E \gets \{\}$, $\text{used} \gets \{\}$ \;
    \For{$v \in V_\text{ranked}$}{
        \eIf{$v$ is leaf and $v \notin \text{used}$}{
            $E \gets E \cup \{v\}$, $\text{used} \gets \text{used} \cup \{v\}$ \;
        }{
            $L \gets \text{leaves}(v) \setminus \text{used}$ \tcp*{Unused leaves in subtree}
            \If{$L \neq \emptyset$}{
                $L_k \gets \text{top\_k}(L, \text{scores}, k)$ \;
                $E \gets E \cup L_k$, $\text{used} \gets \text{used} \cup L_k$ \;
            }
        }
        \If{$|E| \geq K$}{\textbf{break}}
    }
    \Return{$E$}
\end{algorithm}

\section{Experiments}
\subsection{Experimental Setup}
\label{exp:setup}
\paragraph{Datasets and Evaluation Metrics.}
We evaluate our approach on four challenging long document QA datasets:
QASPER \citep{qasper-dataset} for research papers (avg. 4170 words), QuALITY \citep{quality-dataset} for reading comprehension (avg. 5022 words), NarrativeQA \citep{narrativeqa-dataset} for narrative documents (avg. 51,372 words), and MultiFieldQA-zh \cite{bai-etal-2024-longbench} for Chinese documents (avg. 6701 words).
We use F1-Match for QASPER and MultiFieldQA-zh, accuracy for QuALITY, and BLEU(B-1)/ROUGE/METEOR for NarrativeQA.
Token-level F1/Recall evaluate retrieval quality on QASPER.
We fix the retrieved context length across methods for fair comparison.

\paragraph{Baselines.}
We evaluate our approach against several strong retrieval baselines:
\begin{itemize}[leftmargin=10pt, parsep=0pt]
    \item Flatten-chunk splits articles into chunks of maximum 100 words while preserving sentence boundaries for semantic coherence.
    \item Flatten-sentence adopts sentence-level splitting and direct retrieval, providing a direct comparison baseline for our hierarchical approach.
    \item RAPTOR constructs a semantic tree through recursive embedding, clustering, and summarization, with retrieval performed on a collapsed tree structure following \citet{DBLP:conf/iclr/SarthiATKGM24}.
    \item Bisection shares our LLM-enhanced representations and hierarchical retrieval mechanism (Sections~\ref{node_representation} and \ref{Hierarchical_retreival}) but constructs trees by recursively dividing sentences into balanced binary subtrees. This isolates the specific contribution of discourse structure by keeping all other components identical to DISRetrieval.
\end{itemize}

\paragraph{Implementation details.}
We train our language-universal discourse parser on RST-DT \citep{carlson-etal-2001-building} following \citet{yu-etal-2022-rst} with gte-multilingual-base\footnote{huggingface.co/Alibaba-NLP/gte-multilingual-base} backbone \cite{zhang2024mgte}, augmented with GPT-4o-translated Chinese data.
All summarization tasks, including both our approach (as defined in Equation~\ref{eq:LLM_enhance}) and RAPTOR baseline, utilize Llama3.1-8B-Instruct for consistency.
We set threshold $\tau=0$ for QASPER and $\tau=50$ for QuALITY/NarrativeQA (Appendix~\ref{app:tau}).
For sentence encoder (Section~\S\ref{eq:embedding}), we test Sentence-BERT\footnote{multi-qa-mpnet-base-cos-v1} and OpenAI text-embedding-3-large\footnote{https://platform.openai.com/docs/guides/embeddings}.
Top-K is set to 5 (Appendix~\ref{app:ablation_topk}).
Generation models include UnifiedQA-3B, GPT-4.1-mini, and Deepseek-v3.
Details in Appendix~\ref{app:exp_details}.

\subsection{Main Results}
\paragraph{Generation Performance.}
We evaluate our approach across different context lengths (200-400 words) and embedding models (SBERT and OpenAI) on QASPER and QuALITY datasets. Table~\ref{tab:generation} presents the comparative results.
The results reveal three key insights:

\begin{table*}[ht]
\centering
\resizebox{\textwidth}{!}{
\setlength{\tabcolsep}{6pt}
\begin{tabular}{lcccccccccccc}
\toprule
\textbf{Method} & \multicolumn{6}{c}{\textbf{F1-Match (QASPER) / \%}} & \multicolumn{6}{c}{\textbf{Accuracy (QuALITY) / \%}} \\
\cmidrule(lr){2-7} \cmidrule(lr){8-13}
& \textbf{200 (s)} & \textbf{200 (o)} & \textbf{300 (s)} & \textbf{300 (o)} & \textbf{400 (s)} & \textbf{400 (o)}
& \textbf{200 (s)} & \textbf{200 (o)} & \textbf{300 (s)} & \textbf{300 (o)} & \textbf{400 (s)} & \textbf{400 (o)} \\
\midrule
\multicolumn{13}{l}{\textbf{UnifiedQA-3B}} \\
\midrule
\quad flatten-chunk         & 33.97 & 37.50 & 35.41 & 38.46 & 36.13 & 39.03 & 52.97 & 56.28 & 54.46 & 57.53 & 55.23 & 57.62 \\
\quad flatten-sentence      & 35.16 & \underline{38.43} & 37.24 & 39.31 & 37.99 & \underline{40.06} & 53.16 & 56.04 & 54.55 & 56.71 & 55.27 & 57.38 \\
\quad RAPTOR               & 33.57 & 37.46 & 34.95 & \underline{39.96} & 37.00 & 39.53 & 53.60 & 55.99 & 54.75 & 57.00 & 55.51 & 58.53 \\
\quad \textit{Ours} \\
\rowcolor{blue!15}
\quad \quad Bisection      & \underline{36.17} & 37.41 & \underline{37.66} & 39.49 & \underline{38.84} & 39.70 & \underline{55.13} & \underline{57.00} & \underline{56.04} & \underline{58.53} & \underline{57.24} & \underline{60.21} \\
\rowcolor{blue!15}

\quad \quad DISRetrieval  & \textbf{37.36} & \textbf{38.49} & \textbf{39.56} & \textbf{40.03} & \textbf{40.65} & \textbf{40.74} & \textbf{55.56} & \textbf{57.67} & \textbf{57.62} & \textbf{59.64} & \textbf{58.87} & \textbf{60.64} \\

\midrule
\multicolumn{13}{l}{\textbf{GPT-4.1-mini}} \\
\midrule
\quad flatten-chunk         & 37.37 & 41.13 & 41.38 & 43.34 & 42.72 & 44.78 & 60.16 & 65.77 & 63.66 & 69.27 & 67.69 & 71.05 \\
\quad flatten-sentence      & 39.82 & 43.08 & 41.84 & \underline{45.28} & 42.44 & \underline{45.78} & 60.12 & 64.43 & 63.09 & 66.68 & 65.24 & 69.94 \\
\quad RAPTOR               & 37.55 & 40.88 & 39.95 & 43.26 & 42.50 & 43.85 & 61.31 & 64.77 & 63.81 & 67.79 & 67.26 & 70.71 \\
\quad \textit{Ours} \\
\rowcolor{blue!15}
\quad \quad Bisection      & \underline{40.98} & \underline{43.12} & \underline{42.74} & 44.54 & \underline{43.49} & 45.69 & \underline{63.71} & \underline{67.88} & \underline{65.68} & \underline{70.71} & \underline{67.93} & \underline{72.00} \\
\rowcolor{blue!15}
\quad \quad DISRetrieval  & \textbf{40.99} & \textbf{43.45} & \textbf{43.01} & \textbf{45.19} & \textbf{44.95} & \textbf{46.31} & \textbf{64.57}  & \textbf{69.27} & \textbf{66.63} & \textbf{72.44} & \textbf{69.37} &  \textbf{73.54} \\

\midrule
\multicolumn{13}{l}{\textbf{Deepseek-v3}} \\
\midrule
\quad flatten-chunk         & 35.29 & 38.14 & 37.99 & 40.78 & 40.57 & 42.26 & 65.68 & \underline{71.52} & 69.65 & \underline{75.02} & 73.25 & 76.56 \\
\quad flatten-sentence      & 36.51 & \underline{39.27} & 39.41 & 41.57 & 40.82 & 43.04 & 64.14 & 68.74 & 68.41 & 72.24 & 69.65 & 73.63 \\
\quad RAPTOR               & 34.66 & 37.84 & 37.30 & 40.50 & 39.77 & 42.17 & 65.53 & 69.13 & 68.65 & 72.24 & 71.19 & 75.22 \\
\quad \textit{Ours} \\
\rowcolor{blue!15}
\quad \quad Bisection      & \underline{37.28} & \underline{39.27} & \underline{39.80} & \underline{41.81} & \underline{40.96} & \underline{43.57} & \underline{67.83} & 71.09 & \underline{70.81} & 74.40 & \underline{73.39} & \underline{76.94} \\
\rowcolor{blue!15}
\quad \quad DISRetrieval  & \textbf{37.79} & \textbf{39.77} &  \textbf{40.39} & \textbf{42.19} & \textbf{41.51} & \textbf{43.65} & \textbf{68.89} & \textbf{72.28} & \textbf{72.15} & \textbf{76.46} & \textbf{73.68} & \textbf{77.71} \\

\bottomrule
\end{tabular}
}
\caption{
Generation performance comparison of different methods under varying retrieved context lengths (200-400 words) and different embedding models (SBERT: s, OpenAI text-embedding-3-large: o) across two datasets.
Bisection shares our LLM-enhanced node representations and hierarchical retrieval mechanism but uses binary tree construction instead of discourse structure.
Best results are \textbf{bolded}, runners-up are \underline{underlined}.
}
\label{tab:generation}
\end{table*}

\textit{Consistent superiority across settings.}
DISRetrieval outperforms all baselines regardless of context length, embedding model, or generation architecture (UnifiedQA-3B, GPT-4.1-mini, and Deepseek-v3).
For instance, with 400-word contexts and UnifiedQA-3B, we achieve +2.66\% F1-Match on QASPER and +3.60\% accuracy on QuALITY over the flatten-sentence baseline.

\textit{Discourse structure surpasses semantic clustering.}
Compared to RAPTOR's semantic clustering approach, our discourse-aware method demonstrates clear advantages (40.03\% vs. 39.96\% F1-Match on QASPER with OpenAI embeddings), confirming that linguistic discourse structure provides more principled document organization than purely semantic-based methods.

\textit{Linguistic discourse essential for hierarchical modeling.}
The Bisection ablation validates our core hypothesis:
while hierarchical organization helps (Bisection > flatten baselines), incorporating discourse structure is crucial (DISRetrieval > Bisection consistently), providing substantial benefits beyond simple tree-based document modeling.

\paragraph{Retrieval Performance.}
We evaluate DISRetrieval's retrieval effectiveness on QASPER using token-level F1 and Recall metrics.
The Retrieval Result in Table~\ref{tab:retrieval} reveals three key findings:

First, DISRetrieval consistently outperforms baselines across all settings, achieving the highest F1 and Recall scores with both SBERT and OpenAI embeddings.
This demonstrates that our discourse-aware context modeling effectively captures the semantic relationships within documents.

Second, for longer contexts (300-400 words), while all methods show some F1 score degradation, DISRetrieval maintains superior performance, particularly in Recall metrics.
This robust performance on longer contexts validates the capability of our method in handling complex document structures through discourse-guided retrieval.

The ablation with Bisection shows that while hierarchical organization helps, the full discourse-aware approach provides additional benefits.
Additionally, OpenAI embeddings consistently outperform SBERT across all settings, suggesting that strong semantic representations are fundamental to the effectiveness of discourse-aware retrieval.

\paragraph{Cross-Lingual Effectiveness.}

To verify the language-universal capability of our approach, we evaluate on MultiFieldQA-zh, a Chinese long-document QA benchmark.
As shown in Table~\ref{tab:multifieldqa_zh}, DISRetrieval consistently outperforms all baselines across different context lengths with both GPT-4.1-mini (35.25\% F1) and Deepseek-v3 (29.54\% F1).
The improvements are particularly notable at longer contexts, where our method gains +1.30\% to +1.56\% over Bisection with Deepseek-v3.
These results validate that linguistically-grounded discourse structures transcend language boundaries, with our language-universal parser effectively capturing discourse relationships in Chinese documents.
This cross-lingual effectiveness highlights the language-agnostic nature of rhetorical structure theory and demonstrates the extensibility to other languages through similar data augmentation strategies.

\begin{table*}[t]
\centering
\definecolor{mygreen}{RGB}{55,180,9}
\setlength{\tabcolsep}{5pt}
% ---------- Retrieval block (unchanged) ----------
\resizebox{\textwidth}{!}{
\begin{tabular}{lcccccccccccc}
\toprule
\textbf{Retrieval Result}
 &
\multicolumn{2}{c}{\textbf{200 (SBERT)}} & \multicolumn{2}{c}{\textbf{200 (OpenAI)}} &
\multicolumn{2}{c}{\textbf{300 (SBERT)}} & \multicolumn{2}{c}{\textbf{300 (OpenAI)}} &
\multicolumn{2}{c}{\textbf{400 (SBERT)}} & \multicolumn{2}{c}{\textbf{400 (OpenAI)}} \\
% \midrule{2-12}
\cmidrule(lr){2-13}
& \textbf{F1 / \%} & \textbf{Recall / \%} & \textbf{F1 / \%} & \textbf{Recall / \%} 
& \textbf{F1 / \%} & \textbf{Recall / \%} & \textbf{F1 / \%} & \textbf{Recall / \%} 
& \textbf{F1 / \%} & \textbf{Recall / \%} & \textbf{F1 / \%} & \textbf{Recall / \%} \\ 
\midrule
flatten-chunk       & 26.13 & 56.05 & 29.17 & 62.16 & 23.10 & 63.92 & 25.12 & 69.04 & 20.38 & 68.75 & 21.91 & 73.38 \\
flatten-sentence    & 26.15 & 57.80 & 28.25 & 62.78 & 22.68 & 64.63 & 24.04 & 68.43 & 19.98 & 69.02 & 21.06 & 72.42 \\
RAPTOR              & 24.57 & 52.42 & 27.18 & 58.49 & 21.71 & 60.00 & 23.57 & 65.63 & 19.27 & 65.11 & 20.64 & 70.04 \\
\textit{Ours} \\
\rowcolor{blue!15}
 \quad Bisection    & \underline{27.63} & \underline{59.82} & \underline{29.29} & \underline{63.11} & \underline{23.83} & \underline{66.69} & \underline{25.16} & \underline{69.94} & \underline{21.10} & \underline{71.52} & \underline{21.98} & \underline{74.18} \\
\rowcolor{blue!15}
 \quad DISRetrieval & \textbf{28.13} & \textbf{60.75} & \textbf{30.27} & \textbf{65.33} & \textbf{24.62} & \textbf{67.98} & \textbf{26.00} & \textbf{71.71} & \textbf{21.58} & \textbf{72.61} & \textbf{22.79} & \textbf{75.95} \\
\midrule
\end{tabular}
} % end resizebox for retrieval
\vspace{4pt}
%
% ---------- Generation block (横向三大列：Golden | Full | Full Doc Enhanced(6列)) ----------
\resizebox{\textwidth}{!}{
\begin{tabular}{lccccccccc} % 1 label + 1 golden + 1 full + 6 enhanced = 9 cols
\toprule
 \textbf{Generation Result} 
  & \multicolumn{6}{c}{\textbf{Full Document + Retrieved Evidence}}  & \multicolumn{1}{c}{\textbf{Full Document}} & \multicolumn{1}{c}{\textbf{Gold Evidence}}\\
\cmidrule(lr){2-7}
& \textbf{200 (SBERT)} & \textbf{200 (OpenAI)} & \textbf{300 (SBERT)} & \textbf{300 (OpenAI)} & \textbf{400 (SBERT)} & \textbf{400 (OpenAI)}  & \textbf{Avg. 4170 words} & \textbf{Avg. 129 words} \\
\midrule 

\textbf{F1 / \%}
 % & 49.56 & 49.83 & 49.63 & 49.72 & 49.71 & 49.56  & 48.81 & 50.71 \\
  & 49.65 (\textcolor{mygreen}{+0.84})
 & 50.05 (\textcolor{mygreen}{+1.24})
 & 49.77 (\textcolor{mygreen}{+0.96})
 & 50.11 (\textcolor{mygreen}{+1.30})
 & 49.54 (\textcolor{mygreen}{+0.73})
 & 50.20 (\textcolor{mygreen}{+1.39})
 & 48.81
 & 50.71 \textcolor{mygreen}{(+1.90)} \\
\bottomrule
\end{tabular}
} % end resizebox for generation
\caption{
Retrieval and generation results (token-level F1 and Recall) on the QASPER dataset.
Embedding models: SBERT and OpenAI text-embedding-3-large. Generation model: Llama3.1-8B-Instruct.
}
\label{tab:retrieval}
\end{table*}

\begin{table}[ht]
\centering
\setlength{\tabcolsep}{3pt}
\resizebox{\linewidth}{!}{
    \begin{tabular}{lcccccc}
    \toprule
    \textbf{Method} 
    & \multicolumn{3}{c}{\textbf{GPT-4.1-mini (F1 / \%)}} 
    & \multicolumn{3}{c}{\textbf{Deepseek-v3 (F1 / \%)}} \\
    \cmidrule(lr){2-4} \cmidrule(lr){5-7}
    & \textbf{200} & \textbf{300} & \textbf{400}
    & \textbf{200} & \textbf{300} & \textbf{400} \\
    \midrule
    flatten-chunk           & \underline{28.77} & 31.25 & 33.46 & 22.69 & 26.91 & 26.70 \\
    flatten-sentence        & 28.59 & 31.95 & 34.41 & 26.15 & 27.04 & 27.06 \\
    RAPTOR          & 26.61 & 31.70 & 34.32 & 23.11 & 25.96 & 27.01 \\
    \textit{Ours} \\
    \rowcolor{blue!15}
    \quad Bisection     
    & 28.23 
    & \underline{32.31} 
    & \underline{34.80} 
    & \underline{26.71} 
    & \underline{27.05} 
    & \underline{28.24} \\
    \rowcolor{blue!15}
    \quad DISRetrieval  
    & \textbf{29.60} 
    & \textbf{32.97} 
    & \textbf{35.25} 
    & \textbf{26.76} 
    & \textbf{28.61} 
    & \textbf{29.54} \\
    \bottomrule
    \end{tabular}
}
\caption{
Performance on MultiFieldQA-zh (Chinese).
}
\label{tab:multifieldqa_zh}
\end{table}

\begin{table}[ht]
    \centering
    \setlength{\tabcolsep}{3pt}
    \resizebox{\linewidth}{!}{
    \begin{tabular}{lccc}
    \toprule
    & \multicolumn{1}{c}{\textbf{BLEU / \%}} & \multicolumn{1}{c}{\textbf{ROUGE / \%}} & \multicolumn{1}{c}{\textbf{METEOR / \%}} \\
    \midrule
    flatten-chunk   & 24.24 & 28.42 & 19.70  \\
    flatten-sentence    & 21.53 & 28.22 & 18.33  \\
    RAPTOR & \underline{25.05} & \underline{30.24} & \underline{20.92}  \\
    \textit{Ours} \\
    \rowcolor{blue!15}
    \quad Bisection  & 24.71 & 28.85 & 20.41  \\
    \rowcolor{blue!15}
    \quad DISRetrieval & \textbf{25.39} & \textbf{30.31} & \textbf{21.16}  \\
    \bottomrule
    \end{tabular}
    }
    \caption{Performance on the NarrativeQA dataset.}
    \label{tab:narrative}
\end{table}

\subsection{Discussions}

\paragraph{RQ1: Is DISRetrieval effective for handling extremely long documents? Tab~\ref{tab:narrative}}
We evaluate our method on NarrativeQA, which contains exceptionally long documents (average: 51,372 words, maximum: 346,902 words) that exceed most generative models' context limits.
DISRetrieval achieves superior performance across all metrics, outperforming the widely-used flatten-chunk baseline by +1.15\% BLEU, +1.89\% ROUGE, and +1.46\% METEOR.
Notably, both DISRetrieval and RAPTOR substantially outperform flatten-based methods, confirming that structured approaches are more effective than flat for extremely long document retrieval.
However, DISRetrieval's discourse-aware structure provides consistent advantages over RAPTOR's semantic clustering approach.

\paragraph{RQ2: Is precise evidence retrieval essential for effective question answering? Tab~\ref{tab:retrieval}}
To evaluate the importance of precise evidence retrieval, we compare three context settings using Llama3.1-8B-Instruct:
(i) golden evidence, (ii) full document, and (iii) full document augmented with retrieved evidence (highlighted with \textbf{[EVIDENCE] ... [/EVIDENCE]} markers).
Golden evidence achieves the highest F1 score of 50.71\% with only 129 words on average, significantly outperforming the full document approach (48.81\% with 4,170 words).
This demonstrates that concise yet accurate retrieval is more effective than using substantially larger context.
Furthermore, augmenting the full document with our retrieved evidence yields consistent improvements of 0.73-1.39\% across all settings, confirming that precise retrieval enhances performance even when combined with complete document access.
These results underscore that effective long-document QA systems should prioritize retrieval quality over context quantity.

\begin{figure}[ht]
    \centering
    \includegraphics[width=\linewidth]{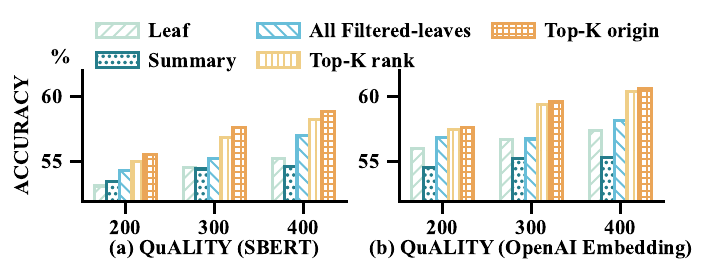}
    \caption{
    Ablation results of different variants.
    }
\label{fig:ablation}
\end{figure}

\paragraph{RQ3: Is the hierarchical retrieval strategy effective? Fig~\ref{fig:ablation}}
We conduct ablation experiments on the QuALITY dataset to evaluate our hierarchical retrieval strategy.
We compare five variants: leaf-only baseline, summary-based retrieval, all filtered-leaves, Top-K with ranking order, and our Top-K with original order.
Three key findings emerge:
(1) Summary-based retrieval underperforms the leaf baseline, confirming that preserving original text details is crucial for effective retrieval.
(2) While all filtered-leaves shows marginal improvements, selective Top-K methods are superior due to reduced noise from irrelevant content.
(3) Our Top-K origin method consistently achieves the best performance by preserving natural document flow.
This advantage intensifies with longer contexts and stronger embeddings, validating that maintaining document structure is essential for hierarchical retrieval.

\paragraph{RQ4: Does the scale of LLMs affect the quality of the discourse-aware tree structures in node text enhancement? Tab~\ref{tab:retrieval_performance}} 
\label{RQ4}
We investigate whether model scale affects discourse-aware tree construction by comparing different LLMs.
As shown in Table~\ref{tab:retrieval_performance}, smaller models like Llama-3.1-8B, Qwen2.5-7B, and Mistral-7B achieve comparable performance to larger models, with differences under 0.5\%.
Notably, Llama-3.1-8B achieves the best recall at 400-word context length.
These results indicate that our approach does not heavily depend on LLM scale, enabling low-cost deployment with 7B models.
Moreover, our method is 3$\times$ faster than RAPTOR (e.g., 50K words: 103s vs. 338s), with preprocessing costs amortized across multiple queries (detailed in Appendix~\ref{comput_efficiency}).

\begin{table}[ht]
    \centering
    \setlength{\tabcolsep}{3pt}
    \resizebox{\linewidth}{!}{
    \begin{tabular}{lcccccc}
    \toprule
    & \multicolumn{2}{c}{\textbf{200}} & \multicolumn{2}{c}{\textbf{300}} & \multicolumn{2}{c}{\textbf{400}} \\
    \cmidrule(r){2-3} \cmidrule(r){4-5} \cmidrule(r){6-7}
    & \textbf{F1 / \%} & \textbf{Recall / \%} & \textbf{F1 / \%} & \textbf{Recall / \%} & \textbf{F1 / \%} & \textbf{Recall / \%} \\
    \midrule
    Llama-3.1-8B   & 28.13 & 60.75 & 24.62 & 67.98 & 21.58 & 72.61 \\
    Qwen2.5-7B    & 28.54 & 61.08 & 24.73 & 68.13 & 21.76 & 72.71 \\
    Mistral-7B & 28.51 & 61.57 & 24.68 & 68.53 & 21.54 & 72.50 \\
    GPT-4o-mini  & 28.29 & 61.63 & 24.65 & 68.42 & 21.49 & 72.65 \\
    Deepseek-v3 & 28.66 & 61.80 & 24.77 & 68.66 & 21.72 & 72.69 \\
    \bottomrule
    \end{tabular}
    }
    \caption{Effect of different LLMs for node text enhancement on QASPER dataset retrieval performance (token-level F1 and Recall) across varying context lengths.}
    \label{tab:retrieval_performance}
\end{table}

\begin{figure}[ht]
    \centering
    \includegraphics[width=\linewidth]{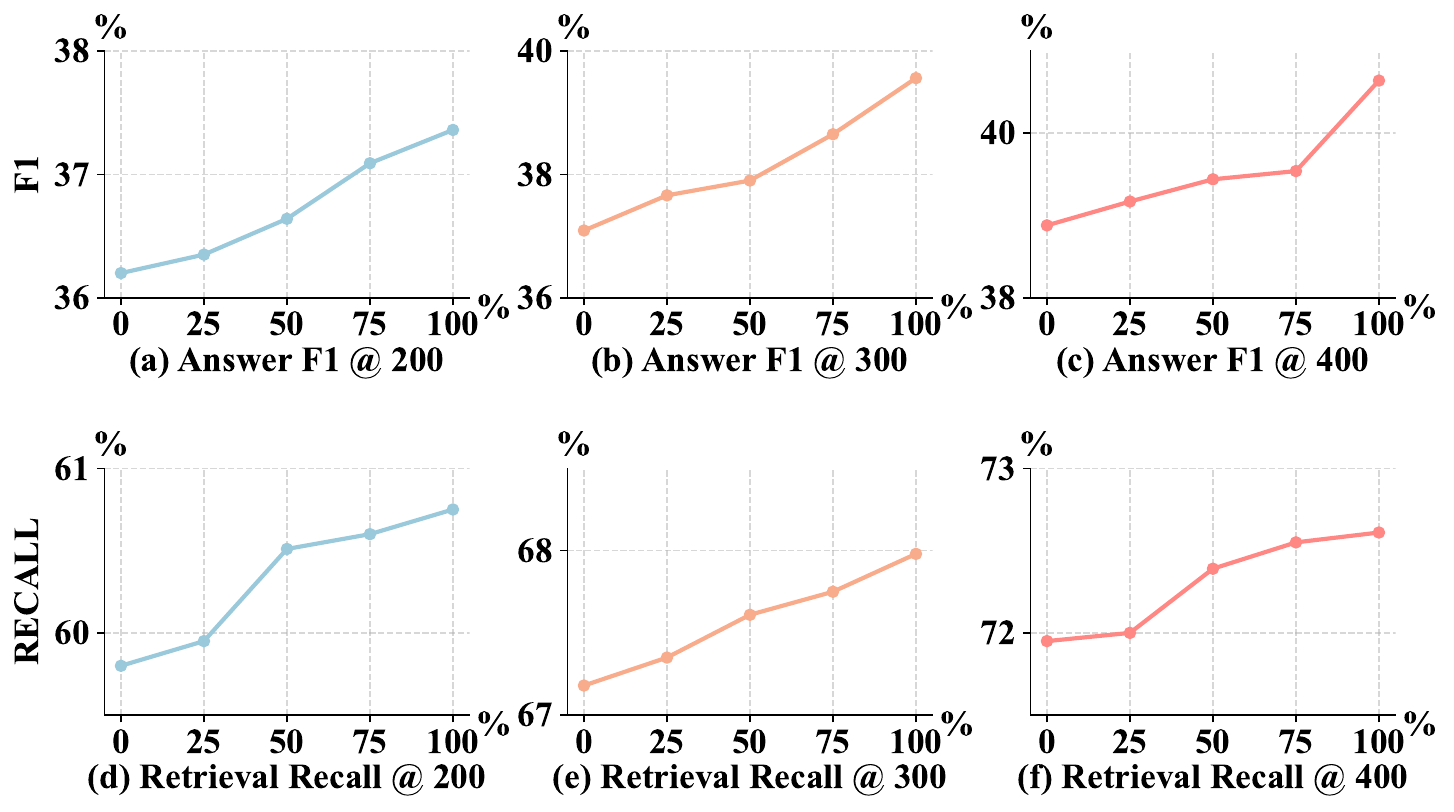}
    \caption{
    Impact of discourse parser capability on subsequent retrieval and question answering performance. Parsers used for comparative evaluation are trained on varying data scales, ranging from 0\% to 100\%.
    }
\label{fig:ablation_parser}
\end{figure}

\paragraph{RQ5: Does the capability of the discourse parser have a significant impact? Fig~\ref{fig:ablation_parser}} 
We examine how discourse parser capabilities affect downstream performance by training parsers on varying amounts of data (0-100\%).
As shown in Figure~\ref{fig:ablation_parser}, both retrieval recall and answer F1 scores improve consistently as parser training data increases across all context lengths.
Notably, despite our parser being trained only on RST-DT (news-text-dominant), DISRetrieval consistently outperforms all baselines across diverse genres (research papers, fiction, movie scripts), demonstrating practical robustness and extensible potential as parsing technology advances.

% \paragraph{RQ6: Does our method work effectively in other languages? Tab~\ref{tab:multifieldqa_zh}}
% To verify cross-lingual effectiveness, we conduct experiments on MultiFieldQA-zh, a Chinese long-document QA benchmark.
% We train a Chinese discourse parser using translated RST-DT by GPT-4o and chinese-xlnet-base\footnote{https://huggingface.co/hfl/chinese-xlnet-base}.
% All other experimental settings are the same as those in QASPER.
% DISRetrieval consistently outperforms all baselines across different context lengths with both GPT-4.1-mini (35.25\% F1) and Deepseek-v3 (29.54\% F1).
% The improvements are particularly notable at longer contexts, where our method gains +0.45\% to +1.30\% over Bisection with Deepseek-v3, validating the language-agnostic nature of leveraging rhetorical structure for long-document retrieval.

\section{Conclusion}

In this paper, we presented DISRetrieval, a discourse-aware hierarchical retrieval framework that systematically incorporates rhetorical structure theory into long document question answering.
Our approach features three key innovations: language-universal discourse parsing, LLM-enhanced node representations, and structure-aware evidence selection.
Comprehensive experiments on QASPER, QuALITY, NarrativeQA, and MultiFieldQA-zh demonstrate substantial improvements over existing methods across varying context lengths, embedding models, generation architectures, and languages.
Ablation studies confirm that discourse structure provides more principled document organization than semantic clustering approaches, with the framework naturally adapting to different document types and information needs through multi-granularity retrieval.
Our cross-lingual experiments validate that linguistically-grounded discourse structures effectively transcend language boundaries, demonstrating robust generalization from English to Chinese documents.
Beyond performance improvements, this work demonstrates that linguistically-grounded approaches remain valuable in the era of LLMs, offering principled alternatives to purely data-driven methods.
Our framework opens new directions for incorporating structured linguistic knowledge into neural retrieval systems, with potential applications in legal document analysis, scientific literature review, and educational content processing.

\section*{Limitations}
While DISRetrieval demonstrates substantial improvements across multiple datasets and architectures, we acknowledge several limitations that point to promising future directions.
First, our discourse parser's performance inherently bounds the overall system effectiveness, as shown in Figure~\ref{fig:ablation_parser}, where parser quality directly impacts downstream performance.
However, our language-universal parser, trained with LLM-based multilingual data augmentation, demonstrates effective cross-lingual generalization from English to Chinese and consistent cross-genre improvements (research papers, fiction, movie scripts), showing practical robustness and extensible potential as parsing technology advances.
While we currently demonstrate effectiveness on English and Chinese, extending to more languages would require additional translated training data through similar augmentation strategies.
We note that the scarcity of suitable long-document QA datasets in other languages currently limits broader multilingual evaluation.
Future work will explore expanding language coverage through collaboration with multilingual NLP communities, developing domain-adaptive parsers, and investigating multi-granularity discourse analysis.
Second, our adaptive summarization strategy using threshold $\tau$ is relatively simple but proves effective across diverse document types ($\tau$=0 for academic papers, $\tau$=50 for narratives).
While this approach successfully balances computational efficiency with representation quality, more sophisticated dynamic thresholding based on content complexity and hierarchical position could further optimize the trade-off.
Third, current evaluation metrics, though comprehensive across four challenging datasets spanning multiple languages, may not fully capture the nuanced benefits of discourse-aware retrieval, such as structural coherence preservation and hierarchical information flow.
We plan to develop evaluation frameworks that better assess discourse-aware retrieval quality across multiple dimensions and languages.

\section*{Ethical Considerations}

This research adheres to ethical principles in natural language processing research and does not raise significant ethical concerns.
Our work focuses on improving document retrieval and question answering systems through discourse-aware hierarchical modeling, which has potential positive societal impacts by enhancing information access and comprehension across multiple languages.
The datasets used in our experiments (QASPER, QuALITY, NarrativeQA, and MultiFieldQA-zh) are publicly available benchmark datasets that have been previously vetted by the research community and do not contain sensitive personal information or harmful content.
Our discourse parsing approach operates on linguistic structures rather than content semantics, minimizing risks of bias amplification or misuse.
The language-universal parser is trained using LLM-based translation for data augmentation, which may inherit potential biases from the translation model, though our focus on structural discourse relations rather than semantic content helps mitigate such risks.
The proposed DISRetrieval framework is designed as a general-purpose retrieval enhancement technique that can benefit various applications requiring long document understanding across languages, such as academic research assistance, educational content processing, and legal document analysis.
We acknowledge that like any information retrieval system, our method could potentially be misused if applied to spread misinformation or manipulate access to information, but such concerns are not specific to our approach and apply broadly to information retrieval technologies.
We encourage responsible deployment of our system with appropriate safeguards and human oversight, particularly in high-stakes applications and cross-lingual scenarios.
All experiments were conducted using publicly available computational resources and open-source tools, ensuring reproducibility and transparency in our research process.

\bibliography{custom}

\appendix

\begin{table*}[t]
\centering
\resizebox{1\linewidth}{!}{
\setlength{\tabcolsep}{8pt}
\begin{tabular}{lccccccccc}
\toprule
\multirow{2}{*}{\textbf{Dataset}} & \multirow{2}{*}{\textbf{Avg. Sentence Length}} & \multicolumn{2}{c}{\textbf{Avg. Mid Node Depth}} & \multicolumn{2}{c}{\textbf{Avg. Leaf Num}} & \multicolumn{2}{c}{\textbf{Avg. Mid Node Percentage / \%}} \\
\cmidrule(lr){3-4} \cmidrule(lr){5-6} \cmidrule(lr){7-8}
 & & \textbf{SBERT} & \textbf{OpenAI} & \textbf{SBERT} & \textbf{OpenAI} & \textbf{SBERT} & \textbf{OpenAI} \\
\midrule
% QASPER  & 22.32 & 5.29 & 5.53 & 12.64 & 13.37 & 54.15 & 65.13 \\
% QuALITY & 14.86 & 13.78 & 17.09 & 67.67 & 88.67 & 73.93 & 88.53 \\
QASPER  & 22.77 & 5.37 & 5.44 & 12.89 & 13.61 & 54.95 & 66.02 \\
QuALITY & 14.64 & 13.98 & 17.34 & 68.84 & 87.92 & 74.86 & 89.95 \\
\bottomrule
\end{tabular}
}
\caption{
Statistical analysis of retrieved intermediate node characteristics across QuALITY and QASPER datasets. The table compares various key metrics: \textit{Avg. Sentence Length} means the average sentence length of all documents; \textit{Avg. Mid Node Depth} is the average depth of retrieved intermediate nodes; \textit{Avg. Leaf Num} represents the average number of leaf nodes that each retrieved intermediate node maps to; \textit{Avg. Mid Node Percentage} is the average percentage of intermediate nodes among the Top-20 retrieved nodes.
}
\label{tab:analysis}
\end{table*}

\section{Extension of Technical Details}
\label{app:technical_details}
In this section, we introduce the specific details of our method which we cannot present in the main article due to the space limit.

\subsection{Discourse Parser Details}
\label{app:discourse_parser}
The discourse parser architecture builds upon a transition-based system that constructs discourse trees through a series of well-defined actions.
This system is particularly effective for capturing both local and global discourse relationships while maintaining computational efficiency.

The transition system maintains two primary data structures: a stack $\sigma$ that holds partially built trees, and a queue $\beta$ that contains unprocessed sentences.
This design follows the intuition that discourse relationships often exist between adjacent text spans and can be built incrementally from bottom to top.

The system builds the tree step by step using three basic operations:
\begin{enumerate}
    \item A ``shift'' action moves a sentence from the queue to the stack when we need new content to process.
    \item A ``reduce'' action combines two adjacent subtrees on top of the stack into a new subtree by identifying their discourse relationship.
    \item A ``pop root'' action concludes the process when we have successfully built a complete tree.
\end{enumerate}
Each state of the system is represented as $c = (\sigma, \beta)$, starting from $c_0 = ([~], S_i)$ with all sentences in the queue, and ending at $c_f = ([T_i], [~])$ with a complete discourse tree $T_i$.

The transition system follows a deterministic process guided by the neural scoring model:
\begin{enumerate}
    \item Initialize $\sigma = [~]$ and $\beta = S_i$.
    \item While $\beta$ is not empty or $|\sigma| > 1$: (a) If $|\sigma| < 2$ and $\beta$ is not empty, perform a ``shift'' action to move the next sentence from $\beta$ to $\sigma$; (b) Else if $\beta$ is empty, perform a ``reduce'' action to combine the top two subtrees in $\sigma$; (c) Else, use the neural scoring model to decide between a ``shift'' or ``reduce'' action based on the current state of $\sigma$ and $\beta$.
    \item Return the single tree $T_i$ remaining on the stack $\sigma$.
\end{enumerate}

The scoring model considers the three topmost subtrees on the stack ($s_1, s_2, s_3$) and the next sentence in the queue $q_1$.
This design is motivated by several factors:
\begin{enumerate}
    \item $s_1$ and $s_2$ are the immediate candidates for the next potential "reduce" action.
    \item $s_3$ provides crucial context about the recently built structure.
    \item $q_1$ helps determine if we should introduce new content via a "shift" action.
\end{enumerate}
For each tree node $v$, we compute its representation $h_v$ recursively:
\begin{equation}
    \mathbf{h}_v = \begin{cases}
    \text{PLM}(s_i), & \text{if } v \text{ is \textit{sentence}} \\
    \frac{1}{|C(v)|}\sum_{u \in C(v)} \mathbf{h}_u, & \text{if } v \text{ is \textit{relationship}}
    \end{cases}
    \label{eq:PLM}
\end{equation}
where $C(v)$ denotes the set of child nodes of $v$, and PLM($\cdot$) is a pre-trained language model that encodes the semantic meaning of individual sentences.
The action scores are then computed as:
\begin{equation}
    \mathbf{y}(a) = \mathbf{W}(\mathbf{h}_{s_1} \oplus \mathbf{h}_{s_2} \oplus \mathbf{h}_{s_3} \oplus \mathbf{h}_{q_1}) + \mathbf{b},
\end{equation}
where $\oplus$ concatenates the representations to capture their interactions, and $\mathbf{W}$ and $\mathbf{b}$ are learnable parameters.
The probability of taking action $a$ is computed using a softmax function over the action scores:
\begin{equation}
    p(a|c) = \frac{\exp(\mathbf{y}(a))}{\sum_{a' \in A} \exp(\mathbf{y}(a'))},
\end{equation}
where $A$ is the set of valid actions at state $c$.
We train the model using supervised learning with gold-standard discourse trees.
The objective function combines cross-entropy loss for action prediction with L2 regularization:
\begin{equation}
    \mathcal{L}(\theta) = -\log p(a^*|c) + \frac{\lambda\|\theta\|_2}{2},
\end{equation}
where $a^*$ is the correct action derived from gold-standard trees.
During inference, we greedily select the highest-scoring action at each step, effectively building the tree in a bottom-up manner while maintaining the discourse relationships between text spans.

\subsection{Iterative Tree Construction Process}
\label{app:iter_tree}

The complete algorithm of our DISRetrieval is presented in Algorithm~\ref{alg:disretrieval}.
Below we detail the iterative tree construction process, which corresponds to Stage 1 of the algorithm.
The construction of discourse trees for long documents presents unique challenges in balancing computational efficiency with structural integrity.
We propose an iterative construction strategy that addresses these challenges through a hierarchical, phase-wise approach.
This method effectively manages computational resources while preserving discourse relationships at multiple granularity levels.
Our iterative process consists of three distinct phases, each designed to handle specific aspects of the tree construction:

\begin{figure*}[t]
    \centering
    \includegraphics[width=\linewidth]{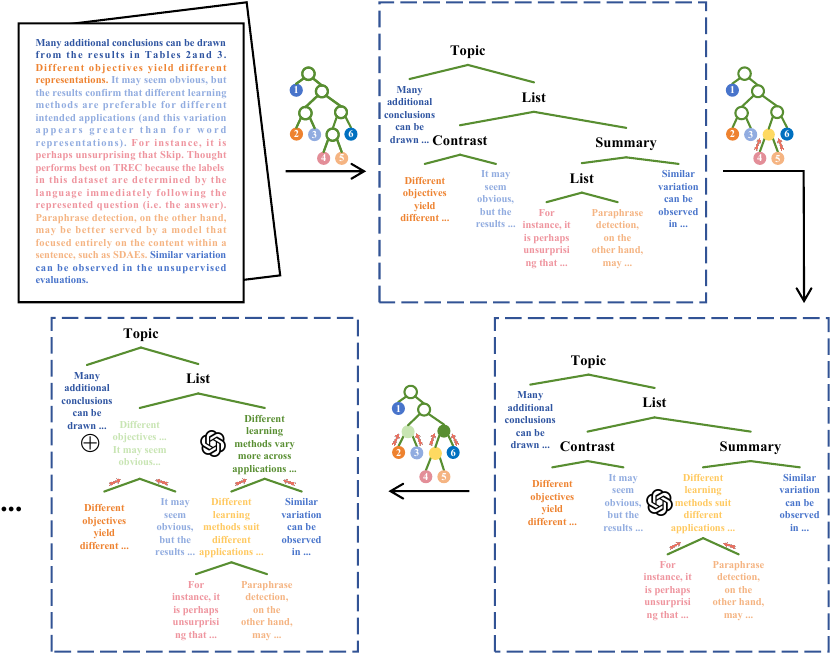}
    \caption{Illustration of bottom-up LLM enhancement in Phase 2 of discourse tree construction. Top: Input paragraph with its initial discourse tree structure. Center: Two-way processing strategy based on text length - using direct concatenation ($\bigoplus$) when combined length is below threshold $\tau$, and LLM-based summarization when above threshold. Bottom: Enhanced discourse tree with progressively generated semantic summaries following Equation \ref{eq:LLM_enhance}, demonstrating the transformation from rhetorical relations to concrete semantic representations.}
    \label{fig:llm_enhancement}
\end{figure*}

\paragraph{Phase 1: Paragraph-level Tree Construction.}
The first phase focuses on building local discourse trees for individual paragraphs. For each paragraph $p_i$ containing sentences $\boldsymbol{S_i} = \{s_{i,1}, s_{i,2}, ..., s_{i,m}\}$, we construct a local discourse tree $T_i$ using our transition-based parsing system. This phase is particularly efficient as it:
\begin{enumerate}
    \item Initializes each paragraph's parsing state with an empty stack and sentence queue: $c_0 = ([~], S_i)$.
    \item Processes paragraphs independently, enabling parallel computation.
    \item Applies transition actions iteratively until a complete tree is formed.
    \item Stores both the resulting paragraph-level tree $T_i$ and its root representation $\mathbf{h}_{T_i}$.
\end{enumerate}

\paragraph{Phase 2: Document-level Tree Construction.}
The second phase focuses on capturing document-level discourse structure. After obtaining all paragraph-level trees $T_1, T_2, ..., T_n$, we:
\begin{enumerate}
    \item For each paragraph-level tree $T_i$, apply bottom-up LLM-enhanced summarization:
    \begin{itemize}
        \item For each non-leaf node $v$ with children $v_l$ and $v_r$:
        \begin{equation}
            v^* = \begin{cases}
                f_\text{LLM}(v_l, v_r), & \text{if } |v_l| + |v_r| \geqslant \tau \\
                f_\text{merge}(v), & \text{otherwise}
                \end{cases}
        \end{equation}
        where $|v_l|$ and $|v_r|$ are the textual content length of child nodes
        \item Continue until reaching root node to obtain semantic unit $u_i$
    \end{itemize}
    \item Form the semantic units set $U = \{u_1, u_2, ..., u_n\}$ from root representations
    \item Apply the discourse parser to these units to construct a document-level tree $T_\text{doc}$ using the same transition-based parsing system:
    \begin{equation}
        T_\text{doc} = f_\text{discourse}(U)
    \end{equation}
    \item Apply bottom-up LLM-enhanced summarization to $T_\text{doc}$:
    \begin{itemize}
        \item For each non-leaf node $v \in T_\text{doc}$ with children $v_l$ and $v_r$:
        \begin{equation}
            t_v = \begin{cases}
                f_\text{LLM}(t_l \oplus t_r), & \text{if } |t_l \oplus t_r| \geqslant \tau \\
                t_l \oplus t_r, & \text{otherwise}
            \end{cases}
        \end{equation}
        \item Process nodes level by level from bottom to top until reaching the root of $T_\text{doc}$
    \end{itemize}
\end{enumerate}

This step effectively captures the high-level discourse relationships between paragraphs while maintaining computational efficiency by working with LLM-enhanced condensed representations at both paragraph and document levels.

\paragraph{Tree Integration.}
The final phase integrates the local and global structures into a unified discourse tree. This integration is accomplished through a careful replacement process:
\begin{equation}
    T_\text{D} = \text{Replace}(T_\text{doc}, \{(u_i, T_i)|i \in [1,n]\})
\end{equation}
where $\text{Replace}(\cdot)$ replaces each unit $u_i$ in the document-level tree with its corresponding paragraph-level tree $T_i$.
This integration phase carefully preserves both local and global discourse relationships by maintaining the internal structure of paragraph-level trees while retaining the document-level relationships established in Phase 2, ultimately creating a seamless hierarchical structure that spans the entire document and effectively captures discourse relationships at all levels of granularity.

The resulting tree structure effectively captures discourse relationships at multiple levels of granularity, from sentence-level connections within paragraphs to broader document-level organizational patterns. This approach not only ensures computational efficiency through its phase-wise processing but also maintains the integrity of discourse relationships throughout the document hierarchy.

\subsection{Detailed LLM Enhancement Process}
\label{appendix:llm_enhancement}

This section provides a detailed illustration of the bottom-up LLM enhancement process described in Phase 2 of our discourse tree construction methodology (Section \ref{section:tree_construction}).

\subsubsection{LLM Enhancement Workflow}

Figure~\ref{fig:llm_enhancement} demonstrates the complete workflow of our LLM enhancement process using a concrete example from a research paper paragraph. The process consists of three main components:

\paragraph{Input Structure.} The top portion shows the input paragraph with its initial sentence-level discourse tree structure. Each sentence is represented as a leaf node, connected through discourse relations such as \textit{Contrast}, \textit{List}, \textit{Summary}, and \textit{Topic}. This hierarchical structure captures the rhetorical organization of the paragraph content.

\paragraph{Adaptive Processing Strategy.} The center portion illustrates our two-way processing strategy based on text length thresholds. When the combined length of child nodes ($|t_l| + |t_r|$) falls below the threshold $\tau$, we apply direct concatenation ($\bigoplus$) to preserve the original textual details. Conversely, when the combined length exceeds $\tau$, we employ LLM-based summarization to generate more concise representations while retaining essential semantic information.

\paragraph{Enhanced Tree Structure.} The bottom portion shows the resulting enhanced discourse tree with progressively generated semantic summaries. Each internal node now contains meaningful textual representations that capture the essence of its subtree content. For example, the \textit{Contrast} relation between different learning methods is transformed into the concrete summary "Different learning methods suit different applications", while maintaining the hierarchical discourse structure.

\subsubsection{Key Benefits}

This LLM enhancement process provides several advantages:
\begin{compactitem}
    \item \textbf{Semantic Preservation}: Maintains essential meaning while reducing textual complexity
    \item \textbf{Computational Efficiency}: Adaptive thresholding prevents unnecessary LLM calls for short text segments
    \item \textbf{Hierarchical Coherence}: Preserves discourse relationships at multiple granularity levels
    \item \textbf{Downstream Compatibility}: Generates representations suitable for sentence-level discourse parsing
\end{compactitem}

The transformation from abstract rhetorical relations to concrete semantic representations enables our framework to effectively bridge the gap between discourse structure and semantic understanding, facilitating improved retrieval performance in long document question answering tasks.

% \begin{figure*}[t]
%     \centering
%     \includegraphics[width=\linewidth]{figures/generation.pdf}
%     \caption{Illustration of bottom-up LLM enhancement in Phase 2 of discourse tree construction. Top: Input paragraph with its initial discourse tree structure. Center: Two-way processing strategy based on text length - using direct concatenation ($\bigoplus$) when combined length is below threshold $\tau$, and LLM-based summarization when above threshold. Bottom: Enhanced discourse tree with progressively generated semantic summaries following Equation \ref{eq:LLM_enhance}, demonstrating the transformation from rhetorical relations to concrete semantic representations.}
%     \label{fig:llm_enhancement}
% \end{figure*}

% \input{tables/algorithm.tex}

\section{Detailed Experiment Settings}
\label{app:exp_details}
\subsection{Dataset Specifications}
The QASPER dataset is constructed from NLP research papers, containing questions that require deep understanding of technical content.
All answerable questions are annotated with multiple reference answers to account for different valid expressions of the same information.
The ground-truth evidence spans are carefully annotated by domain experts to ensure the reliability of retrieval evaluation.

\begin{table}[H]
\centering
\resizebox{1\linewidth}{!}{
\begin{tabular}{lccccc}
\toprule
\textbf{Dataset}   & \textbf{Used Set} & \textbf{Question Num} & \textbf{Doc. Num} & \textbf{Avg. words} & \textbf{Max. words} \\
\midrule
QASPER    & test     & 1456            & 416             & 4170       &          21,165         \\
QuALITY   & dev      & 2086            & 115             & 5022       &          5967       \\
NarrativeQA & test   & 10,577           & 355             &  51,372     &         346,902          \\
MultiFieldQA-zh  &test  & 200             & 200         & 6701         & 14918 \\
\bottomrule
\end{tabular}
}
\caption{
Detailed information of the datasets used in our experiments.
}
\label{tab:datasets}
\end{table}

The QuALITY dataset consists of long passages primarily drawn from fiction stories and magazine articles, with an average length longer than typical QA datasets.
Each question is accompanied by multiple choice options that test comprehensive understanding of the passage.
We utilize the validation set with publicly available labels for our experiments to enable thorough analysis and comparison.
Notably, to enable more extensive experimentation, we evaluated on the validation set with publicly available labels rather than the submission-required test set.

The NarrativeQA dataset is a significant advancement in evaluating machine reading comprehension.
Unlike traditional datasets that focus on short passages, NarrativeQA emphasizes understanding long-form narratives, such as entire books or full-length movie scripts.
One of the defining features of NarrativeQA is the length of its documents.
On average, each document contains 51,372 words.
Notably, some individual documents extend up to 346,902 words, making it one of the longest reading comprehension datasets available.

The MultiFieldQA-zh Chinese dataset comprises long-form documents from various sources, including legal texts, government reports, encyclopedias, and academic papers, each accompanied by relevant questions and answers. Compared to the English documents, the Chinese data exhibits unique linguistic structures, expressions, and domain-specific terminology, posing higher demands on models’ capabilities for long-text comprehension and cross-domain information integration. In our experiments, we use the test set for evaluation and comparison.
More detailed information is presented in Table~\ref{tab:datasets}.

\subsection{RST-DT Data Translation}
\label{RST-DT_Translation}
To train a cross-lingual discorse parser, we translate the RST-DT training data into Chinese to augment the available Chinese data.
Given the syntactic and word-order differences across languages, we perform translation at the sentence level, which is also consistent with our training objective of sentence-level RST parsing.
We then combine the original RST-DT data, the sentence-level RST data, and the Chinese-translated data to form the final training corpus.
For the translation process, we employ the GPT-4o-mini model, and the prompt is provided below:
% \begin{table}[h]
%     \centering
\begin{center}

\definecolor{input_text}{RGB}{180,0,3}

\definecolor{retrieved_text}{RGB}{150,10,200}
\definecolor{llm_text}{RGB}{55,180,9}

\begin{tikzpicture}

    \node[inner sep=0pt] (table) {
        \begin{tabular}{p{0.9\linewidth}}
             \rowcolor{gray!15}
            \textbf{Prompt for Translation}
              \\
            \hline 
            \textbf{System}: You are a helpful assistant.
            \\
            \\
            \textbf{User}: Please translate the following sentences into Chinese. 
            Translate each sentence individually while preserving the original order, 
            and return the results in JSON format. 
            \\
            For example: \\
            Original sentences: \\
            1. Sentence 1 \\
            2. Sentence 2 \\
            3. Sentence 3 \\
            Translation: \\
            \{1: translated sentence, 2: translated sentence, ...\} \\
            \\
            Strictly return the output in valid JSON format. 
            Do not include any additional text, and ensure that the output can be directly parsed as JSON. \\
            \\
            Original sentences: \textcolor{input_text}{\{sentences\}} \\
            \\
            \textbf{LLM}: \textcolor{llm_text}{JSON-formatted Chinese translations.}

        \end{tabular}
    };

    \draw[rounded corners=5pt, line width=1pt] (table.north west) -- (table.north east) -- (table.south east) -- (table.south west) -- cycle;
\end{tikzpicture}
% \end{table}
\end{center}

\subsection{Model Specifications}
For semantic embeddings, the Sentence-BERT model (multi-qa-mpnet-base-cos-v1) is based on the MPNet architecture with 768-dimensional representations, specifically optimized for question-answering tasks.
The OpenAI embedding model (text-embedding-3-large) represents their latest advancement in semantic representation capabilities.

\subsection{Implementation Details}
In the discourse analysis process, we preserve sentence boundaries throughout all splitting operations to maintain semantic coherence.
For the RAPTOR baseline implementation, we follow the original paper's collapsed tree approach where all nodes are considered simultaneously during retrieval.
The tree construction process in our Bisection baseline ensures a nearly balanced binary structure through recursive division of the sentence set.

% \subsection{Evaluation Protocol}
% For QASPER's retrieval evaluation, we compute token-level F1 and Recall scores by comparing the retrieved context against the annotated ground truth evidence spans.
% This provides a fine-grained assessment of retrieval quality beyond simple overlap metrics.
% For QuALITY, the accuracy metric reflects the proportion of correctly answered multiple-choice questions, directly measuring the impact of retrieval quality on downstream QA performance.

\subsection{Experiments Compute Resources}
\label{app:compute_resources}
For the training of the sentence-level Discourse Parser, we used 1 NVIDIA A100-40G GPU.
All other experiments incluing document discourse parsing, LLM summarization and node embedding, as well as the retrieval and generation processes, are conducted using 4 NVIDIA A800-80G GPUs.

\subsection{Prompts}
\definecolor{process_text}{RGB}{180,0,3}
\definecolor{input_text}{RGB}{150,10,200}
\definecolor{llm_text}{RGB}{55,180,9}
We provides the specific prompts used in our approach to ensure reproducibility.

In the prompt templates:
(1) fixed prompts are displayed in black.
(2) Input text is highlighted in \textcolor{process_text}{deep red}.
(3) The retrieved context is colored in \textcolor{input_text}{purple}.
(4) The output generated by the LLM is presented in \textcolor{llm_text}{green}.

% \begin{table}[h]
%     \centering
\begin{center}
\definecolor{input_text}{RGB}{180,0,3}
\definecolor{llm_text}{RGB}{55,180,9}

\begin{tikzpicture}

    \node[inner sep=0pt] (table) {
        \begin{tabular}{p{0.9\linewidth}}
             \rowcolor{gray!15}
            \textbf{Prompt for Intermediate Node Text Summerization}
              \\
            \hline 
            \textbf{System}: You are a helpful assistant.
            \\
            \\
            \textbf{User}: Write a summary of the given sentences, keeps as more key information as possible. Only give the summary without other text. Make sure that the summary no more than 200 words.\\
            Given text: \textcolor{input_text}{\{left child node text\} \{input child node text\}}\\
            \\
            \textbf{LLM}: \textcolor{llm_text}{Summerization result.}
            
        \end{tabular}
    };

    \draw[rounded corners=5pt, line width=1pt] (table.north west) -- (table.north east) -- (table.south east) -- (table.south west) -- cycle;
\end{tikzpicture}
% \end{table}
\end{center}
% \begin{table}[h]
%     \centering
\begin{center}

\definecolor{input_text}{RGB}{180,0,3}

\definecolor{retrieved_text}{RGB}{150,10,200}
\definecolor{llm_text}{RGB}{55,180,9}

\begin{tikzpicture}

    \node[inner sep=0pt] (table) {
        \begin{tabular}{p{0.9\linewidth}}
             \rowcolor{gray!15}
            \textbf{Prompt for Question Answering on QASPER Dataset}
              \\
            \hline 
            \textbf{System}: You are a helpful assistant.
            \\
            \\
            \textbf{User}: Using the following information: \textcolor{retrieved_text}{\{context\}}. 
            Answer the following question in less than 5-7 words, if possible. 
            For yes or no question, only return 'yes' or 'no'. \\
            question: \textcolor{input_text}{\{question\}}\\
            \\
            \textbf{LLM}: \textcolor{llm_text}{Question Answering result.}
            
        \end{tabular}
    };

    \draw[rounded corners=5pt, line width=1pt] (table.north west) -- (table.north east) -- (table.south east) -- (table.south west) -- cycle;
\end{tikzpicture}
% \end{table}
\end{center}
% \begin{table}[h]
%     \centering
\begin{center}

\definecolor{input_text}{RGB}{180,0,3}

\definecolor{retrieved_text}{RGB}{150,10,200}
\definecolor{llm_text}{RGB}{55,180,9}

\begin{tikzpicture}

    \node[inner sep=0pt] (table) {
        \begin{tabular}{p{0.9\linewidth}}
             \rowcolor{gray!15}
            \textbf{Prompt for Question Answering on QuALITY Dataset}
              \\
            \hline 
            \textbf{System}: You are a helpful assistant.
            \\
            \\
            \textbf{User}: Given context: \textcolor{retrieved_text}{\{context\}}.\\
            Answer the following multiple-choice question:\textcolor{input_text}{\{question\}}\\
            \\
            \textbf{LLM}: \textcolor{llm_text}{The correct answer is (A). The context provided...}
            
        \end{tabular}
    };
    
    \draw[rounded corners=5pt, line width=1pt] (table.north west) -- (table.north east) -- (table.south east) -- (table.south west) -- cycle;
\end{tikzpicture}
% \end{table}
\end{center}

\section{Extended Experiment Analysis}

\subsection{Computational Efficiency and Scalability}
\label{comput_efficiency}
\begin{figure}[ht]
    \centering
    \includegraphics[width=\linewidth]{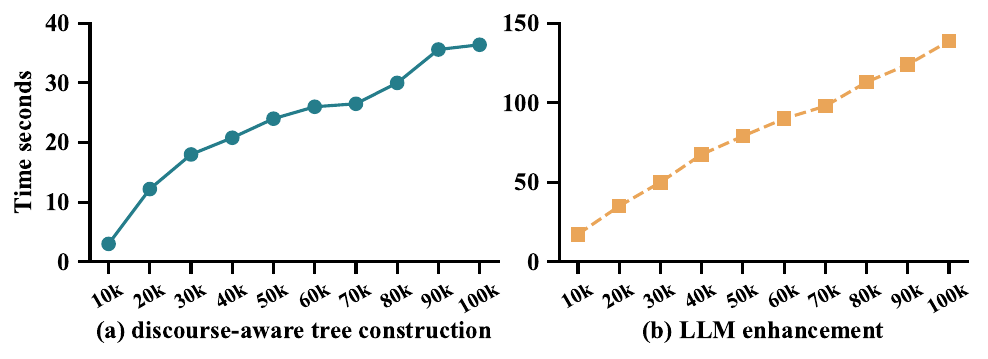}
    \caption{
    Average processing time of discourse-aware tree construction and LLM enhancement across varying document lengths.
    }
    \label{fig:time}
\end{figure}

To evaluate the computational efficiency and scalability of our approach, we conducted comprehensive timing analysis across varying document lengths.
We systematically selected documents ranging from 10,000 to 100,000 words, performed discourse-aware tree construction and LLM enhancement, and measured the computation time for each process.
All experiments were conducted on a single A800 80G GPU, with LLM enhancement performed using the LLaMA3.1-8B-Instruct model accelerated by the vLLM framework.
To ensure statistical reliability, we sampled 10 documents for each length range and report the average processing time with standard deviation.
The detailed results are presented in Figure~\ref{fig:time}.

The processing time for both discourse-aware tree construction and LLM enhancement exhibits near-linear scaling with document length, demonstrating favorable computational complexity for long documents.
Specifically, tree construction for documents up to 100,000 words requires less than 40 seconds, while LLM enhancement completes within 140 seconds, confirming practical feasibility for real-world applications.
The linear scaling behavior indicates that our method maintains consistent per-word processing efficiency regardless of document length.
Moreover, as discussed in Section~\ref{RQ4}, the size and performance of the summarization model have no significant impact on the results, which further suggests that our method can be deployed with a modest computational cost.

Crucially, both tree construction and LLM enhancement are query-independent preprocessing operations performed once per document, allowing costs to be amortized across multiple queries on the same document.
For instance, processing a 50,000-word document requires approximately 90 seconds of preprocessing but can subsequently serve unlimited queries with minimal additional overhead.
In practical deployment scenarios where documents receive multiple queries over time, this one-time preprocessing cost becomes negligible compared to the cumulative benefits of improved retrieval accuracy.

Compared to traditional chunking approaches that require no preprocessing, our method introduces modest upfront costs but delivers significantly better retrieval quality.
The preprocessing overhead is comparable to other hierarchical methods like RAPTOR, while providing superior performance through linguistically-grounded discourse modeling.

\begin{table}[H]
    \centering
    \resizebox{1\linewidth}{!}{
    \setlength{\tabcolsep}{8pt}
    \begin{tabular}{lccccc}
    \toprule
    \textbf{Method} & \textbf{10K words} & \textbf{30K words} & \textbf{50K words} & \textbf{70K words} & \textbf{90K words} \\
    \midrule
    RAPTOR & 63.4s & 194.3s & 338.5s & 488.0s & 611.5s \\
    \rowcolor{blue!15}
    DISRetrieval & \textbf{20.3s} & \textbf{68.1s} & \textbf{103.0s} & \textbf{124.4s} & \textbf{159.9s} \\
    \midrule
    Speedup & 3.1$\times$ & 2.9$\times$ & 3.3$\times$ & 3.9$\times$ & 3.8$\times$ \\
    \bottomrule
    \end{tabular}
    }
    \caption{Computational efficiency comparison between DISRetrieval and RAPTOR using NVIDIA A800 GPU and LLaMA3.1-8B-Instruct.}
    \label{tab:efficiency_comparison}
\end{table}

To provide a comprehensive evaluation of computational efficiency, we conducted a direct comparison between our method and RAPTOR, another preprocessing-based hierarchical retrieval approach.
Both methods were evaluated under identical hardware and software configurations to ensure fair comparison.
Specifically, we used a single NVIDIA A800 GPU with LLaMA3.1-8B-Instruct as the language model for both discourse tree construction in our method and recursive summarization in RAPTOR.

We tested both methods on documents of varying lengths ranging from 10,000 to 90,000 words to assess scalability.
For our method, the total time includes both discourse parsing and LLM-based node enhancement.
For RAPTOR, the time includes the complete recursive summarization process that builds the hierarchical tree structure.
As shown in Table~\ref{tab:efficiency_comparison}, our method demonstrates consistent computational advantages across all document lengths, achieving approximately 3$\times$ speedup compared to RAPTOR.
For instance, processing a 50,000-word document requires 103.0 seconds for our method versus 338.5 seconds for RAPTOR, representing a 3.3$\times$ speedup.
This efficiency advantage becomes even more pronounced for longer documents, with speedup factors reaching 3.9$\times$ for 70,000-word documents.

The computational efficiency of our method stems from two key factors.
First, discourse parsing operates at the sentence level rather than requiring recursive processing of all text segments, which significantly reduces the number of LLM calls needed.
Second, our LLM enhancement process is selective and targeted, focusing only on internal nodes that require summarization, whereas RAPTOR performs recursive summarization across all levels of the hierarchy.
Despite this substantial speedup, our method achieves superior retrieval performance as demonstrated in Tables~\ref{tab:generation}~\ref{tab:retrieval}~\ref{tab:narrative}, indicating that computational efficiency does not come at the cost of effectiveness.

It is important to note that both discourse parsing and LLM enhancement in our framework are query-independent preprocessing steps performed once per document.
In real-world deployment scenarios where multiple queries are issued against the same document collection, these preprocessing costs are amortized across all queries, making the per-query computational overhead negligible.
This characteristic makes our approach particularly suitable for applications such as scientific literature review, legal document analysis, and enterprise knowledge management, where document collections are relatively stable but query patterns are diverse and frequent.

\subsection{Analysis of Multi-granularity Adaptive Retrieval}
We conducted a thorough statistical analysis to evaluate the multi-granularity adaptive retrieval capability of our method.
For each query, we retrieved the Top-20 nodes from the full discourse-aware tree, and analyzed the characteristics of retrieved intermediate nodes across datasets.

\paragraph{Retrieved Node Characteristics Across Datasets.}
Table~\ref{tab:analysis} reveals significant differences in the retrieved intermediate nodes between datasets.
QuALITY exhibits substantially higher values for intermediate node metrics compared to QASPER: the average depth of retrieved intermediate nodes is 2.6 times greater with SBERT embeddings (13.78 vs. 5.29) and 3.1 times greater with OpenAI embeddings (17.09 vs. 5.53).
Similarly, the average number of leaf nodes that each retrieved intermediate node maps to is 5.4 times higher with SBERT (67.67 vs. 12.64) and 6.6 times higher with OpenAI (88.67 vs. 13.37).
The percentage of intermediate nodes among Top-20 retrieved results is also higher for QuALITY across both embedding methods (73.93\% vs. 54.15\% with SBERT; 88.53\% vs. 65.13\% with OpenAI).
These differences exist despite QASPER having a 50\% longer average sentence length (22.32 vs. 14.86 words).

\begin{figure}[ht]
    \centering
    \includegraphics[width=\linewidth]{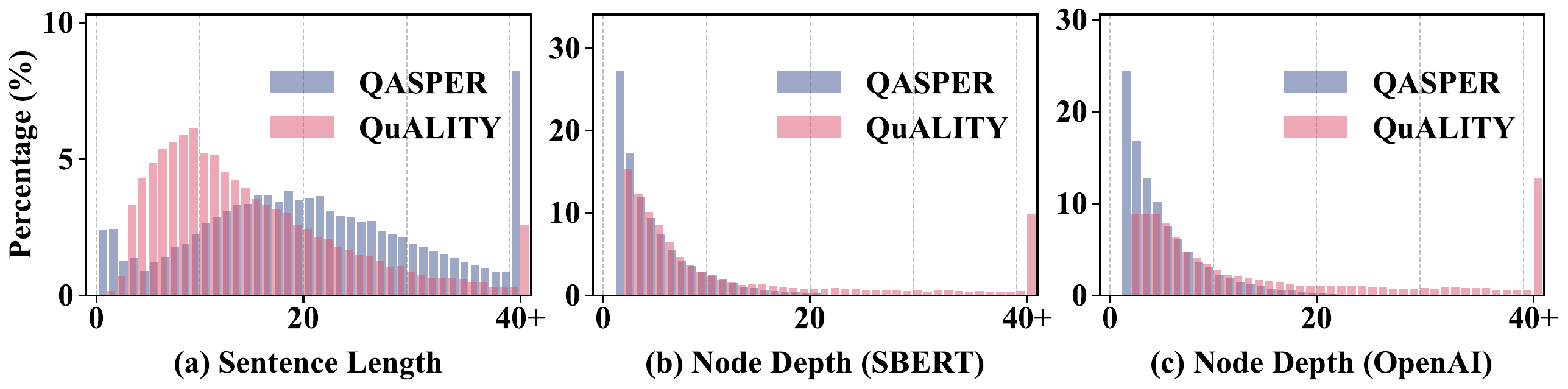}
    \caption{
    Comparative analysis of distribution difference of two datasets.
    Figure (a) shows the difference on Sentence Length, while (b) and (c) demonstrate the distribution of retrieved intermediate node depth across different embedding models.
    }
    \label{fig:distribute}
\end{figure}

\paragraph{Distributional Analysis.}
Figure~\ref{fig:distribute} further illustrates these distinctions through distributional analysis.
The sentence length distribution (Fig.~\ref{fig:distribute}a) shows that QuALITY is heavily skewed toward shorter sentences (5-15 words), while QASPER has a broader, more even distribution extending to 40+ words.
The node depth distributions (Fig.~\ref{fig:distribute}b,c) demonstrate that retrieved intermediate nodes from QuALITY frequently reach depths of 10-15, whereas those from QASPER are predominantly shallower (depths below 10), consistent across both embedding methods.

\paragraph{Retrieval Strategy Adaptation.}
These patterns reveal how our retrieval approach adapts to different document structures.
For narrative fiction in QuALITY with abundant dialogue and shorter sentences, the retrieval system favors higher-level intermediate nodes that aggregate related content, as evidenced by the higher percentage of intermediate nodes and greater node depths.
This suggests that hierarchical composition is particularly beneficial for documents where meaning is distributed across multiple short sentences.
In contrast, for research papers in QASPER with longer, more self-contained sentences, the retrieval system relies less on deep hierarchical structures, as shown by the shallower node depths and lower percentage of intermediate nodes. This indicates that relevant information in scientific documents can often be retrieved effectively with less compositional processing.

These findings empirically validate that our approach effectively adapts to diverse document structures and information needs.
Our discourse-aware tree construction naturally induces appropriate segmentations based on document characteristics rather than arbitrary chunking, enabling multi-level text composition that overcomes the limitations of fixed-granularity methods.
Through these mechanisms, DISRetrieval provides a multi-granularity retrieval framework that adaptively selects appropriate granularity levels across different document types, accommodating their inherent structural variations.

\begin{figure}
    \centering
    \includegraphics[width=\linewidth]{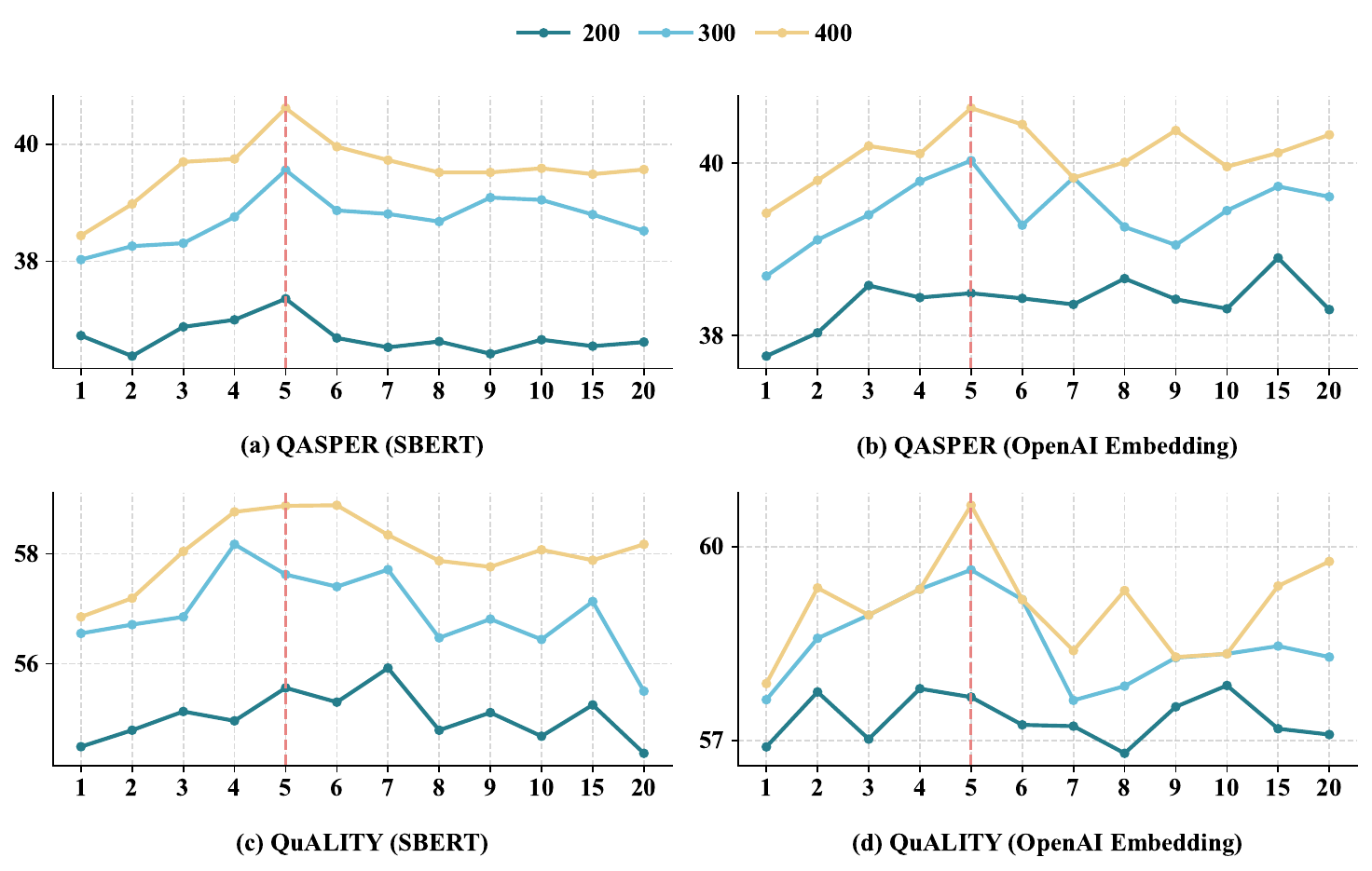}
    \caption{
    Ablation results with different values of $K$.
    The horizontal axis represents different choices of $K$, and the vertical axis indicates generation performance (F1-match for QASPER and accuracy for QuALITY).
    All question answering tasks are conducted on the UnifiedQA-3B model.
    }
    \label{fig:ablation_topk}
\end{figure}

\subsection{Ablation study on different Top-K settings}
\label{app:ablation_topk}
We conduct comprehensive experiments to investigate the impact of varying the value of $K$ from 1 to 20, with results presented in Figure~\ref{fig:ablation_topk}.

Our analysis reveals several noteworthy patterns:

\paragraph{Performance Trends.}
Across all configurations, performance initially improves as $K$ increases, typically peaking around $K=5$ (marked by the vertical dashed line), after which it either plateaus, gradually declines, or exhibits minor fluctuations.
This consistent pattern suggests an optimal balance point where sufficient context is provided without introducing excessive noise.

\paragraph{Dataset-Specific Behaviors.}
QuALITY demonstrates more pronounced performance variations with changing $K$ values compared to QASPER, with performance differences of up to 2 percentage points between optimal and suboptimal settings.
This higher sensitivity likely reflects QuALITY's more complex narrative structure, where precise evidence selection is particularly crucial.

\paragraph{Context Length Impact.}
Longer context windows (300 and 400 words) consistently outperform shorter ones (200 words) across all $K$ values. Notably, the performance advantage of longer contexts is most significant when $K$ is small.
This suggests that when the system retrieves fewer evidence segments, the comprehensiveness of each individual segment becomes critical, as the model must extract all necessary information from a limited number of passages.

Based on these observations, we adopt $K=5$ as our optimal setting for all main experiments, as it consistently delivers strong performance across datasets and embedding methods while maintaining computational efficiency.

\begin{figure}[ht]
    \centering
    \includegraphics[width=\linewidth]{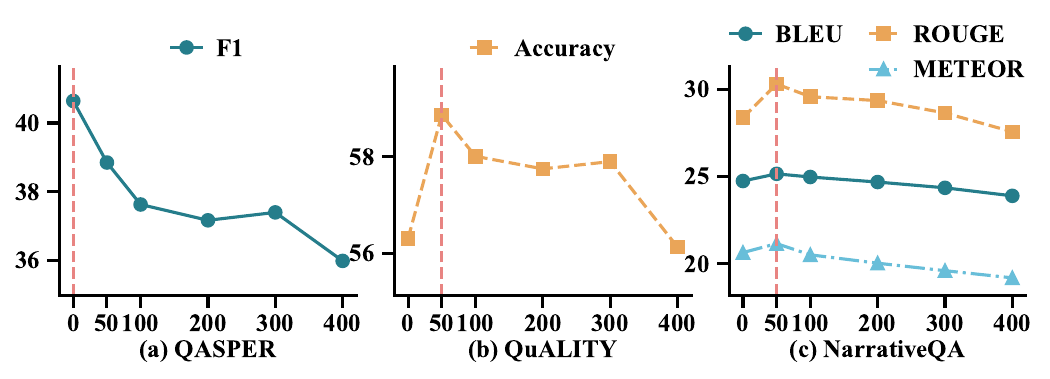}
    \caption{
    Ablation results with different values of $\tau$ across three datasets. Sentence-BERT is used as the embedding model and UnifiedQA-3B as the question-answering model, with a context length of 400 words.
    }
    \label{fig:ablation_tau}
\end{figure}

\subsection{Ablation study on parameter \texorpdfstring{$\tau$}{tau} across different datasets}
\label{app:tau}
Equation~\ref{eq:LLM_enhance} employs a threshold parameter $\tau$ to decide whether use an LLM for summarization or directly merge the subtree.
To assess the impact of this parameter and identify the optimal settings, we conduct experiments on three datasets.
The results are shown in figure~\ref{fig:ablation_tau}.

\paragraph{Performance Trends.}
For QASPER, the F1 score is highest when $\tau=0$ and gradually declines as $\tau$ increases, indicating that summarization is more beneficial than direct merging sentences of subtrees.
% In contrast, the performance on QuALITY and NarrativeQA improves when $\tau=50$ and decreases afterwards, suggesting that a moderate threshold balances summarization and merging effectively. 
In contrast, QuALITY and NarrativeQA reach their peak performance at $\tau=50$ and gradually decline afterward, only approaching the level observed at $\tau=0$ when $\tau$ reaches 300 and 200 respectively, suggesting that a moderate merging strategy is essential for these datasets.

\paragraph{Dataset Analysis}
These results are consistent with the characteristics of the datasets.
QASPER consists of academic papers, where sentences are relatively independent and logically complete.
So keeping sentences as the minimal unit allows for more precise information extraction. 
In contrast, QuALITY and NarrativeQA mainly contain novels and movie scripts, which include numerous dialogues and short sentences that require a relatively complete context to convey the correct semantics.
For these two datasets, merging subtrees based on rhetorical structure parsing results enables logically related sentences to form more coherent text segments, preventing semantic fragmentation.
At the same time, moderate merging is crucial, as excessively long text segments may introduce redundancy.
Therefore, we set $\tau=0$ for QASPER, $\tau=50$ for QuALITY and NarrativeQA respectively. This can also provide guidance for other datasets: for documents with more formal and complete sentences, smaller $\tau$ values are preferable, whereas for documents with shorter and more fragmented sentences, relatively larger $\tau$ values should be used.

\begin{algorithm*}[t]
    \caption{DISRetrieval: Discourse Structure-based Long Document Retrieval}
    \label{alg:disretrieval}
    
    \KwIn{Document $D$ containing sentences $S_i$, paragraphs $P_i$; Query $q$}
    \KwOut{Retrieved evidence segments $E$}

    \tcc{Stage 1: Discourse-Aware Tree Construction}
        \For{each paragraph $P_i \in D$}{
            Initialize stack $\sigma$ and sentence queue $\beta \gets S_i$\;
            \While{$\beta$ not empty OR $|\sigma| > 1$}{
                \tcc{RST parsing operations}
                $a_t \gets$ Select action from $\{$shift, reduce, pop\_root$\}$\tcp*{Determine next parsing action}
                \Switch{$a_t$}{
                    \Case{shift}{
                        Move next sentence from $\beta$ to $\sigma$\tcp*{Add new sentence to stack}
                    }
                    \Case{reduce}{
                        $s_2, s_1 \gets$ Pop top two elements from $\sigma$\;
                        Determine discourse relation $r$ between $s_1$ and $s_2$\tcp*{Using RST relations}
                        Create new node with relation $r$\;
                        Push new node to $\sigma$\;
                    }
                    \Case{pop\_root}{
                        Final tree node $\gets$ Pop from $\sigma$\tcp*{Complete the tree}
                    }
                }
            }
            $T_i \gets$ Resulting discourse tree\;
            
            \tcc{LLM Enhancement for Paragraph Tree}
            \For{each non-leaf node $v \in T_i$ (bottom-up order)}{
                % $t_l, t_r \gets$ Get content from left and right children\;
                $v_l, v_r \gets$ Get left and right children\;
                % \eIf{$|t_l| + |t_r| \geq \tau$}{
                \eIf{$ |v_l| + |v_r| \geq \tau $} {
                    % $t_v \gets f_\text{LLM}(t_l, t_r)$\tcp*{Generate concise summary using LLM}
                    
                    $v^* \gets f_\text{LLM}(v_l, v_r)$\tcp*{Generate concise summary using LLM}
                }{
                    % $t_v \gets t_l \oplus t_r$\tcp*{Direct concatenation for short text}
                    $v^* \gets f_{merge}(v)$\tcp*{Merge the subtree rooted at v }
                }
            }
            Store root text representation $t^i_\text{root}$ for $T_i$\;
        }

        Initialize stack $\sigma$ and queue $\beta \gets \{t^1_\text{root}, ..., t^n_\text{root}\}$\;
        \While{$\beta$ not empty OR $|\sigma| > 1$}{
            \tcc{Similar RST parsing at document level}
            Apply RST parsing operations as in Phase 1\;
            Build document tree $T_\text{doc}$\;
        }
        
        \tcc{Stage 2: Node Representation}
        \For{each non-leaf node $v \in T_\text{doc}$ (bottom-up order)}{
            $t_l, t_r \gets$ Get content from left and right children\;
            \eIf{$|t_l| + |t_r| \geq \tau$}{
                $t_v \gets f_\text{LLM}(t_l, t_r)$\tcp*{LLM-based enhancement for document-level nodes}
            }{
                $t_v \gets t_l \oplus t_r$\tcp*{Direct concatenation for short text, don't need to merge subtree}
            }
        }
        $T_D \gets$ Replace leaf nodes in $T_\text{doc}$ with corresponding $T_i$\tcp*{Integrate trees into unified structure}

        \For{each node $v \in T_D$}{
            $e_v \gets \text{Encoder}(t_v)$\tcp*{Generate dense vector representations}
        }
    
    \tcc{Stage 3: Hierarchical Evidence Retrieval and Selection}
    $e_q \gets \text{Encoder}(q)$\tcp*{Transform query to embedding space}
    $\text{scores} \gets \{\}$, $E \gets \{\}$\;
    \For{node $v \in T_D$}{
        $\text{scores}[v] \gets \text{cosine}(e_q, e_v)$\tcp*{Compute relevance scores}
    }

    Apply Algorithm \ref{alg:retrieval} for evidence selection\;
    
    \Return{$E$}
\end{algorithm*}

% \subsection{Detailed Results of different answer types on QASPER Dataset}
% The question-answering task on the QASPER dataset includes four types of answers:  Abstractive, Yes / No, Extractive and Answerable / Unanswerable.

% \begin{itemize}
%     \item Abstractive: This type of answers use abstract and summarized content rather than the original text from the document. 
%     They often involve rephrasing or synthesizing content to provide a concise and coherent response.

%     \item Yes / No: These answers are straightforward and binary, simply \textit{Yes} or \textit{No} to the question.
    
%     \item Extractive: These answers are directly taken from the original text.
%     The goal is to pinpoint and extract the specific segment of text that directly answers the question.

%     \item Answerable / Unanswerable: This type categorizes whether the question can be answered based on the information provided in the text. 
%     These type of answers correspond to questions that the original documents do not contain sufficient supporting information, and the model is expected to directly reply with "unanswerable."
% \end{itemize}

% We present the detailed results for each type of answer in Table~\ref{tab:answer_type}.
% Notably, both GPT-4.1-mini and Deepseek-v3 achieve zero scores on the Answerable / Unanswerable type due to their inability to meet the specific answer word requirement.

% \input{tables/qasper_answer_type}

% This is an appendix.

\end{document}